\documentclass[useAMS,usenatbib,usegraphicx,onecolumn]{mn2e}

\newcommand{\be}{\begin{equation}}
\newcommand{\ee}{\end{equation}}

\newcommand{\lsim}{\stackrel{<}{\sim}}
\newcommand{\bea}{\begin{eqnarray}}
\newcommand{\eea}{\end{eqnarray}}
\newcommand{\bean}{\begin{eqnarray*}}
\newcommand{\eean}{\end{eqnarray*}}
\newcommand{\bx}{{\bf x}}
\newcommand{\bk}{{\bf k}}

\newcommand{\ba}{\begin{eqnarray}}
\newcommand{\ea}{\end{eqnarray}}
\newcommand{\mat}{{\bf}}
\newcommand{\nn}{\nonumber\\}

\def\nicefrac#1/#2{\leavevmode\kern.1em
\raise.5ex\hbox{\the\scriptfont0 #1}\kern-.1em
/\kern-.15em\lower.25ex\hbox{\the\scriptfont0 #2}}

\title[Multiple Methods for Estimating the Bispectrum...]{Multiple Methods for Estimating
the Bispectrum of the Cosmic Microwave Background with Application to the MAXIMA Data}

\author[M.~G.~Santos et al.]{
  M.~G.~Santos,$^{1,2}$\thanks{\tt santos@thphys.ox.ac.uk}
  A.~Heavens,$^{11}$
  A.~Balbi,$^{4,5,6}$
  J.~Borrill,$^{7,9}$
  P.~G.~Ferreira,$^{1}$
  \newauthor
  S.~Hanany,$^{8,5}$
  A.~H.~Jaffe,$^{13}$
  A.~T.~Lee,$^{10,5,6}$
  B.~Rabii,$^{5,10}$
  P.~L.~Richards,$^{10,5}$
  \newauthor
  G.~F.~Smoot,$^{10,5,6,9}$
  R.~Stompor,$^{7,9,3}$
  C.~D.~Winant$^{5,10}$
  and J.~H.~P.~Wu$^{12}$\\
\vspace*{6pt}
\\$^1$ Astrophysics \& Theoretical Physics, University of Oxford,
Oxford OX1 3RH, UK
\\$^{2}$ 
Dept. of Physics, One Shields Ave., University of California, Davis,
CA 95616, USA
\\$^3$
Dept. of Astronomy, 601 Campbell Hall, University of California,
  Berkeley, CA94720-3411, USA
\\$^4$
Dipartimento di Fisica, Universit\`a Tor Vergata, Roma,
Via della Ricerca Scientifica, I-00133,
  Roma, Italy
\\$^5$
Center for Particle Astrophysics, 301 Le Conte Hall, University of
  California, Berkeley, CA94720-7304, USA
\\$^6$
Lawrence Berkeley National Laboratory, 1 Cyclotron Road,
 Berkeley, CA94720, USA
\\$^7$
National Energy Research Scientific Computing Center,
  Lawrence Berkeley National Laboratory, Berkeley, CA94720, USA
\\$^8$
School of Physics and Astronomy, 116 Church St. S.E., University of
  Minnesota, Minneapolis, MN55455, USA
\\$^9$
Space Sciences Laboratory, University of California,
  Berkeley, CA94720, USA
\\$^{10}$
Dept. of Physics, University of California,
  Berkeley, CA94720-7300, USA
\\$^{11}$ Institute for Astronomy, University of Edinburgh,
Blackford Hill, Edinburgh EH9 3HJ, U.K.
\\$^{12}$ Dept. of Physics, National Taiwan University,
No. 1 Sec. 4 Roosevelt Road, Taipei 106, Taiwan
\\$^{13}$
Astrophysics Group, Blackett Lab., Prince Consort Road,
Imperial College, London SW7 2BW, UK}

\begin{document}

\maketitle

\begin{abstract}
 We describe different methods for estimating the bispectrum of
Cosmic Microwave Background data. In particular we construct
a minimum variance estimator for the flat-sky limit and compare
results with previously-studied frequentist methods.  
Application to the MAXIMA dataset shows consistency
with primordial Gaussianity.  Weak quadratic non-Gaussianity is
characterised by a tunable parameter $f_{NL}$, corresponding to
non-Gaussianity at a level $\sim 10^{-5}f_{NL}$ (ratio of non-Gaussian
to Gaussian terms), and we find limits of
$|f_{NL}|<950$ for the minimum-variance estimator and $|f_{NL}|<1650$
for the usual frequentist estimator.  These are the tightest limits on
primordial non-Gaussianity which include the full effects of the
radiation transfer function.
 \end{abstract}


\section{Introduction}
\label{Introduction}

A primary aim of contemporary cosmology is to explain the origin of 
structure in the universe. The current standard cosmological
model has been able to explain a number of critical observations
such as the recession of galaxies,
the abundance of light elements and the relic, black-body radiation.
A key assumption is that the universe is uniform on large scales.
We know however that the distribution of matter is not homogeneous
and it is therefore necessary to construct extensions to the
standard model which can explain these irregularities. 

All theories for the inhomogeneity of space and matter rely on
mechanisms which generate perturbations of a stochastic nature.
The theory of inflation relies on the amplification and conversion
 of small scale quantum fluctuations into classical fluctuations
of space time on large scales. Alternatively, theories of topological
defects lead to the generation of classical inhomogeneities through
thermal phase transitions in the early universe; in this case the
stochasticity is due to the thermal fluctuations in the very early 
universe
and the highly non-linear evolution of the topological defects. 
Other alternatives have been proposed in the recent years, based
on the assumption that we live in a three-dimensional membrane
in a higher dimensional space-time. The interactions and collisions
 between various ``branes'' in this cosmos will lead to fluctuations
in our space time. It must therefore be clear that to
identify the correct theory of structure formation one must
attempt to characterize its stochastic properties. This must
be done in terms of our observable universe. 

An excellent observable for undertaking such an analysis
is the Cosmic Microwave Background (CMB). Photons from the
time of recombination will have an imprint of fluctuations
in the density of radiation and baryons at that time. These
fluctuations are sufficiently small that one can consider them
as linear perturbations on a homogeneous space-time. Hence one
should be able to characterize the stochastic nature of 
inhomogeneities without it having been obscured by 
subsequent non-linear gravitational collapse. Measurements
of the CMB have already led to a major breakthrough 
in tackling the problem of structure formation. The COBE 
satellite showed that the variance of fluctuations on large
scales is consistent with the principle of cosmological
scale-invariance \citep{COBE}: the variance of fluctuations in the gravitational
potential (a generalization of the Newtonian potential) at early
time is the same on all scales. Recent measurements from
the TOCO \citep{TOCO}, MAXIMA \citep{hanany}, BOOMERanG \citep{BOOM}, 
DASI \citep{DASI}, VSA \citep{VSA} and CBI \citep{CBI} experiments, of the variance
of fluctuations on smaller scales have revealed a preferred
scale of 1$^{\rm o}$ consistent with adiabatic initial fluctuations
in a Euclidean space. It is believed that in the next few years
an accurate picture of the fluctuations in the CMB down to
sub-arcminute scales will emerge.

Given the quality of the data it is timely to undertake a more
complete characterization of the statistical properties of 
inhomogeneities through an analysis of the CMB data.  In addition,
primordial non-Gaussianity should be easier to detect in the CMB than
in large-scale structure \citep{VWHK00}.  
In all cases of interest, the Probability Distribution Function (PDF) 
can be completely described
in terms of its spatial n-point correlation functions, which are 
the expectation
values of all possible products of the random field with itself at
different points in space. Under the assumption of statistical
isotropy and homogeneity,
it is normally more useful to characterize
the PDF in terms of higher order moments of the Fourier
transform of the field. Most readers are familiar with the 2-point
moment, the power spectrum of fluctuations ($C_\ell$). Indeed current
efforts in the analysis of CMB data
have focused mainly on increasingly precise estimates of the
angular power spectrum. For perturbations induced by Inflation,
which is the current favourite model of structure formation, the
CMB power spectrum plays a central role, since the statistics
are very close to Gaussian and all non-zero moments of order 
$n>2$ can be expressed in terms of the $C_\ell$. 

In this paper we will present a detailed analysis of the bispectrum.
The bispectrum is the cubic moment of the Fourier transform
of the temperature field and it can be seen as a scale-dependent 
decomposition of the skewness of the fluctuations (in
much the same way as the $C_\ell$ is a scale dependent decomposition
of the variance of fluctuations). The bispectrum can be used
to look for the presence of a non-Gaussian signal in the CMB sky.
We use the 
data collected with the MAXIMA-1 experiment \citep{hanany} to quantify the 
bispectrum of the CMB.
The Gaussianity of this data set has already been analysed using
complementary methods in \citet{wu01}, including the methods of 
moments, cumulants,
the Kolmogorov test, the $\chi^2$ test, and Minkowski functionals
in eigen, real, Wiener-filtered and signal-whitened spaces and
with one particular estimator of the bispectrum \citep{santos}.
In the past few years, interest in the bispectrum has grown in the
scientific community. Estimates of the bispectrum
in the COBE data proved the statistics to be extremely sensitive to some
non-Gaussian
features in the data, be they cosmological or systematic \citep{Hea98};
the quality of galaxy surveys has made it possible to test for the
hypothesis that the matter overdensity is a result of non-linear 
gravitational collapse of Gaussian initial conditions \citep{SCO01,FD01}. 
On the other
hand a serious effort has been undertaken to calculate the expected
bispectrum from various cosmological effects; secondary anisotropies,
such as the Ostriker-Vishniac effect, lensing, Sunyaev-Zel'dovich
effect \citep{SG99,GS99,CH00}, as well as primordial sources, such as non-linear 
corrections
to inflationary perturbations or cosmic seeds, may lead to observable
signatures in the bispectrum of the CMB \citep{luo94b,v00,j94,gm00b,bcm99,
cm01,gpw01,WK00,KWS02,KS01}. 

In this paper we study some possible estimators for the bispectrum
and use them to check for Gaussianity of the MAXIMA CMB map and to
constrain possible theoretical models of the bispectrum.
After a few definitions, we discuss in section \ref{mle} the minimum variance
cubic estimator and develop its application
in the the small sky approximation. We then use this estimator
to analyse the MAXIMA data and test for the Gaussian hypothesis.
In section \ref{freq} we review the standard frequentist estimator and
apply it to the same dataset so as to compare with the 
previous method. Finally in section \ref{cosmo} we discuss
the weak non-linear coupling model as a specific source
for a primordial bispectrum and use the previously obtained
values to constrain the free parameter in this theory.

\section{Formalism and Notation}
\label{notation}

The bispectrum is defined to be the harmonic transform of the 3-point 
correlation function:
\begin{equation}
B_{\ell_1\ell_2\ell_3}^{m_1 m_2 m_3} \equiv \langle
a_{\ell_1 m_1} a_{\ell_2 m_2} a_{\ell_3 m_3} \rangle,
\label{Blm}
\end{equation}
where the angle brackets indicate ensemble averages and the CMB temperature 
fluctuation field, $\Delta T(\Omega)$ was
expanded into spherical harmonics
\begin{equation}
\label{alm}
a_{\ell m} \equiv \int\,{\rm d}\Omega\, \Delta T(\Omega)
Y_{\ell m}^*(\Omega).
\end{equation}
$\Omega$ corresponds to the spherical polar coordinates $(\theta,\phi)$ and
${\rm d}\Omega$ represents an element of solid angle.
Rotational invariance of the Universe together with invariance under reflections means
we can write the bispectrum in the following form (see \citealt{H01}):
\begin{equation}
  \label{eq:func}
  B_{\ell_1\ell_2\ell_3}^{m_1m_2m_3}
  ={\cal G}_{\ell_1\ell_2\ell_3}^{m_1m_2m_3}b_{\ell_1\ell_2\ell_3}, 
\end{equation}
where $b_{\ell_1\ell_2\ell_3}$ is a real symmetric function of $\ell_1, \ell_2, \ell_3$ and
${\cal G}_{\ell_1\ell_2\ell_3}^{m_1m_2m_3}$ is the Gaunt integral defined by
\bea
\lefteqn{{\cal G}_{\ell_1\ell_2\ell_3}^{m_1m_2m_3}
\equiv
\int {\rm d}\Omega\,
Y_{\ell_1 m_1}(\Omega)
Y_{\ell_2 m_2}(\Omega)
Y_{\ell_3 m_3}(\Omega)}
\nn & &
=\sqrt{\frac{(2\ell_1+1)(2\ell_2+1)(2\ell_3+1)}{4\pi}}
\left({\ell_1\ \ell_2\
\ell_3}\atop{0\ \, 0\ \, 0}\right)
\left({\ell_1\ \ell_2\
\ell_3}\atop{m_1\ m_2\ m_3}\right).
\label{gauntint}
\eea
${\cal G}_{\ell_1\ell_2\ell_3}^{m_1m_2m_3}$ is a purely geometrical term and will impose all
the well known selection rules for the bispectrum (e.g. \citealt{Edmonds}, \citealt{Luo94})
 - it will be zero unless
\begin{enumerate}
\item $\ell_1+\ell_2+\ell_3={\rm even}$
\item $m_1+m_2+m_3=0$
\item $|\ell_i-\ell_j|\le \ell_k \le \ell_i+\ell_j {\rm \ for\ }i,j,k=1,2,3$.
\end{enumerate}
All the relevant cosmological information will be contained in $b_{\ell_1\ell_2\ell_3}$, 
called the reduced bispectrum \citep{KS01}, and this is the quantity we 
try to measure experimentally so as to compare with possible theoretical predictions.
In the literature it is also common to use the angle-averaged bispectrum, related
to the reduced bispectrum by
\bea
\label{angle_average}
\lefteqn{B_{\ell_1\ell_2\ell_3}\equiv\sum_{m_1 m_2 m_3}
\left({\ell_1\ \ell_2\
\ell_3}\atop{m_1\ m_2\ m_3}\right)
B_{\ell_1\ell_2\ell_3}^{m_1m_2m_3}}\nonumber\\ & &
=\sqrt{\frac{(2\ell_1+1)(2\ell_2+1)(2\ell_3+1)}{4\pi}}
\left({\ell_1\ \ell_2\
\ell_3}\atop{0\ \, 0\ \, 0}\right)
b_{\ell_1\ell_2\ell_3}.
\eea
In the case of the MAXIMA-1 experiment \citep{hanany}, the patch analysed is nearly 
flat, and thus it will be more natural to work in the small sky 
approximation where a map of the CMB can be considered approximately flat \citep{White}.
We can then expand $\Delta T(\bx)$ in terms of the 2-dimensional Fourier
modes (after the flat-sky conversion $\Omega \rightarrow \bx$),
\be
\label{fourierdef}
\Delta T(\bx)=\int\frac{{\rm d}^2k}{(2\pi)^2}a(\bk)
e^{i\bk\cdot\bx}.
\ee
The Fourier transform of the 3-point correlation function will be given in this
case, by
\be
\label{bspdef}
\langle a(\bk_1)a(\bk_2)a(\bk_3)\rangle=(2\pi)^2B({k_1},
{k_2},{k_3})\delta^2(\bk_1+\bk_2+\bk_3),
\ee
where the 2D delta function $\delta^2$ 
is a consequence of assuming translational invariance
for the 2-dimensional surface. Again $B({k_1},{k_2},{k_3})$ is a real function,
invariant under permutations of ${k_1},{k_2},{k_3}$. Invariance under rotations
and reflections on the plane means $B({k_1},{k_2},{k_3})$ will only depend on the moduli of 
the $\bk$ vectors. On small angular scales there is a simple correspondence
between the flat sky bispectrum and the full sky angular bispectrum,
\begin{equation}
 \label{flat_approx}
  b_{\ell_1\ell_2\ell_3}\approx
  B(k_1,k_2,k_3),
\end{equation}
with $|\bk_i|=\ell_i$ (see Appendix \ref{bsp_flat}). 
Therefore, in the flat-sky approximation, there is no need to keep 
track of the Wigner 3-j symbols.
Throughout the paper when we refer to the
bispectrum in the flat-sky limit we will mean the above definition, 
that will correspond, to a good
approximation, to the so-called reduced bispectrum.

\section{An Approximate Minimum Variance Estimator}
\label{mle}

\subsection{General setup}
\label{mle_gen}

We seek an estimator of the reduced bispectrum, 
$b_{\ell_1\ell_2\ell_3}$, for an arbitrary set of
pixelized temperature measurements in a CMB map, with Gaussian
noise with arbitrary correlations. Ideally, the estimator should
be unbiased, and optimal in the sense that its error is as small
as possible. If possible, it should also be lossless, so that it contains
as much information as the original CMB map and have
calculable statistical properties.
We want to obtain an expression for the minimum variance estimator
without assuming anything specific to the map being used. We
start with a set of measurements $\{\Delta T_i\}$, making the
pixelized map ($\Delta T_i$ is the temperature fluctuation 
in a sky pixel labeled by $i$). The actual shape of the map,
or whether we use a full sky or ``flat-sky'' analysis, will
be irrelevant for this derivation.

As in \citet{Hea98}, and in the spirit
of the optimal quadratic estimator for the power spectrum in
\citet{Teg97b}, we seek an estimator for the bispectrum which is
cubic in the $\Delta T_i$. The actual derivation of this estimator
is reviewed in Appendix \ref{cub_gen} in the context of a full-sky 
analysis. Here we just show the main
steps to recover the minimum variance cubic estimator.
We start by considering quantities $y_\alpha$ of the following form
\begin{equation}
y_\alpha = \sum_{{\rm pixels}\ ijk} E^\alpha_{ijk}\ \Delta T_i 
\Delta T_j \Delta T_k,
\label{y_est}
\end{equation}
where the $E^\alpha_{ijk}$ are weights to be determined, and
$\alpha$ is a shorthand for the triplet $\{\ell_1,\ell_2,\ell_3\}$.
The $y_\alpha$ are related to the bispectrum
estimates, but will not be the bispectrum estimates themselves.

The mean of $y_\alpha$ involves the 3-point correlation function,
which in turn, should be related to the reduced bispectrum, 
$b_\alpha$ as follows:
\begin{equation}
\langle \Delta T_i \Delta T_j \Delta T_k\rangle =
\sum_\alpha b_\alpha Q_{ijk}^\alpha
\label{3-point}
\end{equation}
The expression for $Q_{ijk}^\alpha$ depends on the actual
basis being used to decompose the temperature measurements.
It is defined
in the appendix (eq. \ref{Qijk}), for the full sky setup, but 
is not important here.
The ensemble average of $y$ is then
\begin{equation}
\langle y_\alpha \rangle = \sum_{\alpha'} b_{\alpha'}
F_{\alpha \alpha'},
\end{equation}
where 
\begin{equation}
\label{fisher_def}
F_{\alpha \alpha'} = \sum_{ijk} Q^{\alpha'}_{ijk} E^{\alpha}_{ijk}.
\end{equation}
$F_{\alpha \alpha'}$ is the Fisher and covariance matrix of 
the quantity $y_\alpha$.
We obtain the
weights for the optimal estimator of the reduced
bispectrum, by minimizing the error on $y_\alpha$ (e.g.
$\langle y_\alpha y_{\alpha'}\rangle$).  In order to effect this, we
assume that the field is close to Gaussian, and approximate the
6-point correlation function by that for a Gaussian field (see
Appendix B).  The optimal weights are
\begin{equation}
E^\alpha_{ijk} = {1\over
6}\xi_{ii'}^{-1}\left[\xi_{jj'}^{-1}\xi_{kk'}^{-1}
 - {3\over 2+3n}\,\xi_{jk}^{-1}\xi_{j'k'}^{-1}\right]
Q_{i'j'k'}^\alpha
\label{E}
\end{equation}
where the summation convention is assumed and $n$ is the number
of pixels used. This is obtained by considering that the probability
distribution function for the temperature field
is close to a Gaussian. $\xi_{ij}$ is then the
2-point correlation function of this temperature
field, $\xi_{ij} \equiv \langle \Delta T_i \Delta T_j\rangle$.
Finally, the estimator for the reduced bispectrum will be
\begin{equation}
\hat b_\alpha = F_{\alpha \alpha'}^{-1}y_{\alpha'}.
\label{b_y}
\end{equation}
Hence to obtain the bispectrum, we just need
to calculate $Q_{ijk}^\alpha$, whose expression is 
defined by equation (\ref{3-point}). We then 
obtain $\xi_{ij}$ by using the already-measured 
power spectrum and
plug all this into equation (\ref{E}). Finally, we obtain $y_\alpha$ using
equation (\ref{y_est}) which can, at least in principle, be
inverted to get the bispectrum (eq. \ref{b_y}).
The derivation obtained here is symbolic and does not
depend specifically on the characteristics of the map.
In the flat-sky approximation it will be possible to 
obtain an expression equivalent to equation (\ref{3-point})
and then all the above derivation will follow through
in the same way.

\subsection{The flat case}
\label{flat}

In the previous section we discussed what should be the minimum
variance cubic estimator
for the bispectrum in a general setup. This estimator could in principle
be applied to the MAXIMA dataset by expanding the temperature map into
spherical harmonics and calculating $Q_{ijk}^\alpha$ in equation 
\ref{3-point} (see appendix \ref{flat_estimat}). 
Since we will be looking at a small patch of the sky,
it is possible to use the flat-sky approximation, and we will now derive
the expression for the best cubic estimator using this approximation.

The temperature measured by MAXIMA ($\Delta T_s(\bx)$) can be expressed in 
the following way:
\be
\Delta T_s(\bx)=W(\bx)\left[\int {\rm d}^2x'\Delta T(\bx'){\cal B}(|\bx-\bx'|)
+\Delta T_N(\bx)\right],
\ee
where ${\cal B}(x)$ represents the effect of the beam smearing
and pixelisation and $W(\bx)$ is the window function defining 
the field of observation. 
$\Delta T(\bx)$ is the ``true'' temperature of the sky (the one we
would like to measure) and
$\Delta T_N(\bx)$ is the temperature fluctuation due to the noise.

Using the Fourier convention in equation (\ref{fourierdef}), the measured
Fourier modes are related to the underlying, ``true'' temperature ones, by
\be
\label{fmode}
a_s(\bk)=\int\frac{d^2k'}{(2\pi)^2}\,\tilde{W}(\bk-\bk')\left[
a({\bk'})\,\tilde{\cal B}(k')+a_N({\bk'})\right].
\ee
This expression shows clearly the problems associated with the extraction of 
information from the temperature map: there will be a contamination
due to the noise and beam. Even if we ignore these contributions, the $a(\bk)$
will necessarily be convolved by the window function and we can never hope
to measure a single mode without full sky coverage.

To obtain the estimator one first needs to calculate $Q_{ijk}^\alpha$
as it appears in equation (\ref{3-point}). The precise steps for the derivation
of the estimator in the flat-sky approximation can be found in appendix
\ref{flat_estimat}. Here we just present the main expressions.
Assuming the noise bispectrum is zero, we can express
the 3-point correlation function as
\bea
\lefteqn{\langle \Delta T_s(\bx_i) \Delta T_s(\bx_j)
\Delta T_s(\bx_k) \rangle =
\int_{0}^{\infty}d\ell_1\int_{\frac{\ell_1}{2}}^{\ell_1}
d\ell_2\int_{\ell_1-\ell_2}^{\ell_2}d\ell_3\,B(\ell_1,\ell_2,\ell_3)}
\nn & & \times
\tilde{\cal B}(\ell_1)\tilde{\cal B}(\ell_2)\tilde{\cal B}(\ell_3)
\sum_{{\rm perm.}\atop\{\ell_1,\ell_2,\ell_3\}}\int_{0}^{\pi}d\theta_1\,
{\cal R}e\{R_{ijk}(\ell_1,\ell_2,\ell_3,\theta_1)\}
\eea
Where the sum is over all the possible permutations of $\ell_1$, $\ell_2$
and $\ell_3$. The triangle relations are automatically imposed by the limits of 
integration ($\bx_i$ defines the position of pixel $i$ in two dimensions).
$R_{ijk}(\ell_1,\ell_2,\ell_3,\theta_1)$ will basically correspond
to a product of exponentials, for each pixel index, of the type
$e^{i\,\bk\cdot\bx}$, where $|\bk|=\ell$ (eq. \ref{ap:Rdef} and 
eq. \ref{ap:Gdef}). This expression
should be compared to equation \ref{3-point}. The sum for each multipole was replaced by an
integral as expected. We can now read $Q_{ijk}^\alpha$ from the above 
expression (approximating the integral by a sum) and
following the previous discussion, the expression for $y_\alpha$ becomes:
\bea
\label{flat_ydef}
\lefteqn{y(\ell_1,\ell_2,\ell_3) = \tilde{\cal B}(\ell_1)\tilde{\cal B}(\ell_2)
\tilde{\cal B}(\ell_3)\,
{1\over 6}\sum_{{\rm perm.}\atop\{\ell_1,\ell_2,\ell_3\}}\int_{0}^{\pi}d\theta_1
{\cal R}e\{R_{i'j'k'}(\ell_1,\ell_2,\ell_3,\theta_1)} \nonumber \\ & &
\times\xi_{ii'}^{-1}\left[
\xi_{jj'}^{-1}\xi_{kk'}^{-1}
 - {3\over 2+3n}\,\xi_{jk}^{-1}\xi_{j'k'}^{-1}\right]\,
\Delta T_s(\bx_i)\Delta T_s(\bx_j)\Delta T_s(\bx_k)\}
\eea
The fact that the matrix $R_{ijk}(\ell_1,\ell_2,\ell_3,\theta_1)$ can be decomposed 
into a product of objects each with just one pixel index,
represents a major simplification in computational terms.
Then, the sum
over the pixels for $R_{ijk}$ in equation (\ref{flat_ydef})
can be decomposed into just a product of three operations of 
order $n$. 
What is this expression actually doing? To understand it, let us concentrate on the integrand
of $\theta_1$ and first assume the pixel values
are uncorrelated (i.e. $\xi_{ij}^{-1}\approx \frac{\delta_{ij}}{\left<\Delta T^2\right>}$).
Then the first term, involving 
$\xi_{ii'}^{-1}\xi_{jj'}^{-1}\xi_{kk'}^{-1}$,
is just what one would do naively, i.e. discrete Fourier transform (DFT) of the
data for each of the three wave vectors (whose sum is zero) and multiply the coefficients together.
The second term (with the ${3 \over 2+3n}$ factor) would be a product of one DFT of the data with
a DFT of the window function (basically $a_s(\bk)\,\tilde{W}({\bk})$). This term should be small
for ${\bk}$s larger than the fundamental mode (as defined by the size of the field of
 observation).
As discussed later on, this argument has been backed up with some numerical results and for the 
modes 
we analyzed from the MAXIMA data we can safely ignore this second term.
Looking closer at the first term, and now considering the full covariance matrix, we see that
it will basically correspond to a product of three DFTs not of the map itself, but of 
$z_i=\xi_{ij}^{-1}\Delta T({\bx_j})$. This will then be integrated along $\theta_1$ so that
the result is rotationally invariant. The advantages of analyzing the map $z_i$ instead of the
original
map, have already been discussed in \cite{Teg97b}. 
Weighting the pixels by the inverse of the covariance
matrix will suppress the large scale power contribution and soften the edges of the map. This 
should help to reduce the ``red leakage'' from lower to higher multipoles and decrease the width of
the window function.

Estimating $y$ is not however the end of the story: we need to relate
it to the underlying ``true'' bispectrum. The connection is simply
\be
\label{ytobsp}
\langle y(\ell_1,\ell_2,\ell_3)\rangle=
\int_{0}^{\infty}d\ell_1'\int_{\frac{\ell_1'}{2}}^{\ell_1'}d\ell_2'
\int_{\ell_1'-\ell_2'}^{\ell_2'}d\ell_3'
\,F(\ell_1,\ell_2,\ell_3,\ell_1',\ell_2',\ell_3')\,B(\ell_1',\ell_2',\ell_3'),
\ee
where, assuming the almost Gaussian approximation,
\be
F(\ell_1,\ell_2,\ell_3,\ell_1',\ell_2',\ell_3')=
\langle y(\ell_1,\ell_2,\ell_3)\,y(\ell_1',\ell_2',\ell_3')\rangle.
\ee
Having all this information, we can now try to estimate the bispectrum. The next section will
describe several of the numerical details involved when applying these techniques to the 
MAXIMA dataset.

\subsubsection{Numerical Implementations}
\label{numimp}

We are interested in obtaining as many bispectrum values as computationally possible. 
The first task in estimating the bispectrum is to calculate $y$.
Numerical implementation of equation (\ref{flat_ydef}) is straightforward. However, the second 
term in the integration of $\theta_1$, 
involving ${3\over 2+3n}\,\xi_{jk}^{-1}\xi_{j'k'}^{-1}\xi_{ii'}^{-1}$,
will mix pixels from $R_{ijk}(\ell_1,\ell_2,\ell_3,\theta_1)$.
Then the sum over the pixels will effectively be of order $n^2$, which can make the
process quite slow. Fortunately, as previously 
discussed, for high enough multipoles this term should be negligible. The MAXIMA map used has a
fundamental mode of $\Delta\ell \approx 40$ and we are typically interested in modes for 
$\ell > 110$. Numerical tests also show that the contribution of the second term is always less
than $0.1\%$ and so we choose to neglect this term.

To estimate the covariance matrix of $y$, we could use directly equation (\ref{fisher_def})
(see also eq. \ref{ap:fisher_def}). However,
this proves to be somewhat slow, and we decided to use instead Monte-Carlo simulations to
quantify the covariance. After obtaining $y$, we need to relate it to the bispectrum
(eq. \ref{ytobsp}). 
We see the relation requires in principle the knowledge of 
a 6-dimensional object, which is certainly not an easy task. Things become more involved
here than when analysing the 
power spectrum. The straightforward way would be to discretize the integral 
and invert this relation as in equation (\ref{b_y}).
Then the estimator would be unbiased: $\left<\hat{B}(\ell_1,\ell_2,\ell_3)\right
>={B(\ell_1,\ell_2,\ell_3)}$. However 
this would have problems, not only
because it requires the calculation of the Fisher matrix for a large number of points,
which is time 
consuming, but also because inverting $F_{\alpha\alpha'}$ 
is numerically quite difficult since the Fisher
matrix is almost singular. Instead we used the following estimator for the bispectrum:
\be
\hat{B}(\ell_1,\ell_2,\ell_3)=\frac{y(\ell_1,\ell_2,\ell_3)}
{M(\ell_1,\ell_2,\ell_3)},
\ee
where
\be
\label{M}
M(\ell_1,\ell_2,\ell_3) = \int_{0}^{\infty}d\ell_1'\int_{\frac{\ell_1'}{2}}^{\ell_1'}d\ell_2'
\int_{\ell_1'-\ell_2'}^{\ell_2'}d\ell_3'
\,F(\ell_1,\ell_2,\ell_3,\ell_1',\ell_2',\ell_3').
\ee
This normalizes the Fisher matrix, so that
\be
\label{ytobsp_norm}
\langle \hat{B}(\ell_1,\ell_2,\ell_3)\rangle=
\int_{0}^{\infty}d\ell_1'\int_{\frac{\ell_1'}{2}}^{\ell_1'}d\ell_2'
\int_{\ell_1'-\ell_2'}^{\ell_2'}d\ell_3'
\,w(\ell_1,\ell_2,\ell_3,\ell_1',\ell_2',\ell_3')\,B(\ell_1',\ell_2',\ell_3'),
\ee
with
\be
\label{window}
w(\ell_1,\ell_2,\ell_3,\ell_1',\ell_2',\ell_3')\equiv
\frac{F(\ell_1,\ell_2,\ell_3,\ell_1',\ell_2',\ell_3')}
{M(\ell_1,\ell_2,\ell_3)}.
\ee
$w$ can be considered as the measured bispectrum window function, and satisfies
\be
\int_{0}^{\infty}d\ell_1'\int_{\frac{\ell_1'}{2}}^{\ell_1'}d\ell_2'
\int_{\ell_1'-\ell_2'}^{\ell_2'}d\ell_3'
\,w(\ell_1,\ell_2,\ell_3,\ell_1',\ell_2',\ell_3')=1.
\ee
The larger the patch of the sky analyzed, the more narrow this window should be and 
eventually one approaches
\be
\langle\hat{B}(\ell_1,\ell_2,\ell_3)\rangle\approx 
B(\ell_1,\ell_2,\ell_3).
\ee
Otherwise, the estimated bispectrum will just correspond to a weighted average
of the target bispectrum.
Calculating $M(\ell_1,\ell_2,\ell_3)$ would require in principle the sampling of
the Fisher matrix into a fine enough grid so we can obtain a good approximation to
the integral in equation (\ref{M}). This is exactly what we would like to avoid.
Instead, taking into account the expression for the Fisher matrix (eq.
\ref{ap:fisher_def}), it is possible to simplify the integral:
\bea
\label{M_use}
\lefteqn{M(\ell_1,\ell_2,\ell_3)=
{1\over 6}\int d^2x \,{\cal B}(|\bx_{i'}-\bx|){\cal B}(|\bx_{j'}-\bx|)
{\cal B}(|\bx_{k'}-\bx|)}
\nn & &
\times\tilde{\cal B}(\ell_1)\tilde{\cal B}(\ell_2)\tilde{\cal B}(\ell_3)
\,\sum_{{\rm perm.}\atop\{\ell_1,\ell_2,\ell_3\}}\int_{0}^{\pi}d\theta_1
{\cal R}e\left\{R_{i'j'k'}(\ell_1,\ell_2,\ell_3,\theta_1)\right\}\,\xi_{ii'}^{-1}
\xi_{jj'}^{-1}\xi_{kk'}^{-1}.
\eea
The integration over the beam functions can be easily done if we approximate
them by a Gaussian. We then obtain an expression that is easy to deal with
numerically (eq. \ref{ap:M_G}). The fit of the MAXIMA beam and pixel 
window function to a Gaussian might have
some problems on the tails of the function. However we have checked that this
difference only introduces small corrections to the calculation of 
$M(\ell_1,\ell_2,\ell_3)$ and can be ignored. Note that $M(\ell_1,\ell_2,\ell_3)$
does not depend directly on the temperature map $\{\Delta T_i\}$ and therefore
we only need to calculate it once.

\subsubsection{The data}
\label{mldata}

\begin{figure}
\begin{center}
\includegraphics[width=0.6\textwidth]{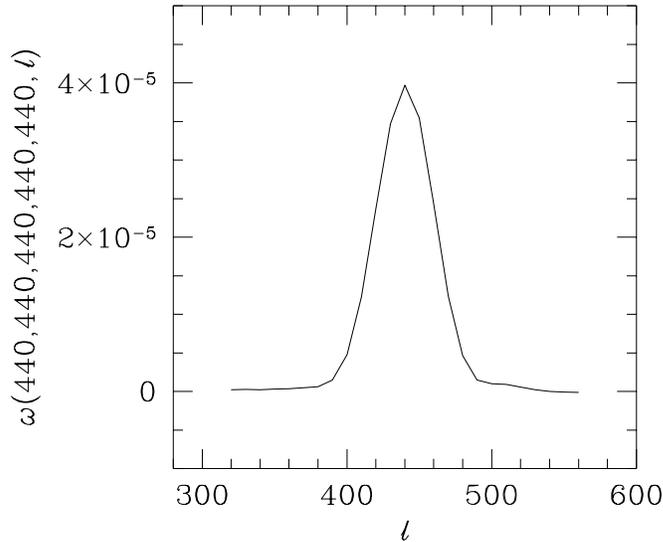}
\vspace{-1.5cm}
\end{center}
\caption{Normalized window function for $\hat{B}(\ell_1,\ell_2,\ell_3)$
 ($\ell=320,330,...,560$).}
\label{fig_window}
\end{figure}

\begin{figure}
\begin{center}
\includegraphics[width=0.6\textwidth]{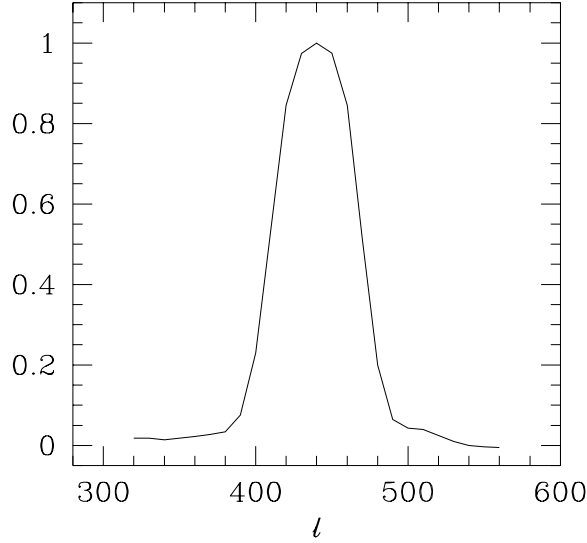}
\vspace{-1.5cm}
\end{center}
\caption{Correlation between $\hat{B}(440,440,440)$ and 
$\hat{B}(440,440,\ell)$. Multipole
values are the same as in Fig. \ref{fig_window}. The rounded peak is due to the sharp
increase of the variance for the diagonal term, $\hat{B}(440,440,440)$.}
\label{fig_correl}
\end{figure}

Using the above simplifications, we can now try to estimate the bispectrum from the MAXIMA
data. The MAXIMA-1 experiment and dataset is described in detail in \citet{hanany}.
As in the previous paper (\citealt{santos}) we used a map with square pixels of 8' each but
considered a slightly larger area.
We used the central region of the map which is $60\,deg^2$ and contains
$\approx 3300$ pixels. This patch has the most uniform sampling and has the highest
signal-to-noise ratio. We estimated the bispectrum from $\ell=110$ up to $\ell=740$ which
corresponds approximately to the same multipole range as in the previous analysis.
For the size of the steps along this multipole range, in 
principle one could choose $\Delta\ell=1$. However, not only this is very
demanding in computational terms, but also we do not gain any extra information
in making the bins so fine, since there will be strong correlations between the
different measurements of order $2\pi/\theta$, where $\theta$ is the linear size
of the smallest dimension of the sky patch (\citealt{bjkpspec}, 
\citealt{Teg97b}).
Although the region analyzed is not exactly rectangular, it is possible to 
get an estimate of the fundamental frequency mode using the above expression,
which gives $\Delta\ell\approx 40$. This can be confirmed by looking at 
Fig.~\ref{fig_window} were we show the window function (eq. \ref{window}) for
a few multipole values. As expected, the width of the window is approximately 
twice this fundamental mode. Figure~\ref{fig_correl}
gives an indication of the level of correlations between different measurements of 
the bispectrum.
We have chosen a step of $\Delta\ell=30$ which will make it easier later on when
comparing to theory. We can always combine the values into larger bins to 
decrease the correlations.

\subsubsection{Results}
\label{mlresults}

Taking into account all the bispectrum symmetries,
our choice of multipole range and step size, gives a total
of 1409 bispectrum estimates.

To measure the covariance matrix, $<\hat{B}(\ell_1,\ell_2,\ell_3)
\hat{B}(\ell'_1,\ell'_2,\ell'_3)>$, we use Monte-carlo (MC) simulations
of the MAXIMA-1 dataset. In the near-Gaussian approximation we are considering,
the variance of the bispectrum is determined by the input power spectrum.
To generate the MC maps we used a Gaussian signal with the power spectrum
from the best fit model obtained in \citet{balbi}: $\Omega_{cdm}=0.6,
\Omega_b=0.1,\Omega_\Lambda=0.3,n=1.08$ and $h=0.53$ (normalized to the MAXIMA-1
and COBE/DMR data). We also took into account the effects
of pixelization and finite beam and included the full noise correlation matrix from
the MAXIMA-1 experiment.
We applied exactly the same estimator to 4000 of these simulated maps, obtaining
1409 modes for each realization. The values for the covariance matrix
actually converge quite rapidly, so that 4000 maps should be enough.

\begin{figure}
\begin{center}
\includegraphics[width=0.6\textwidth]{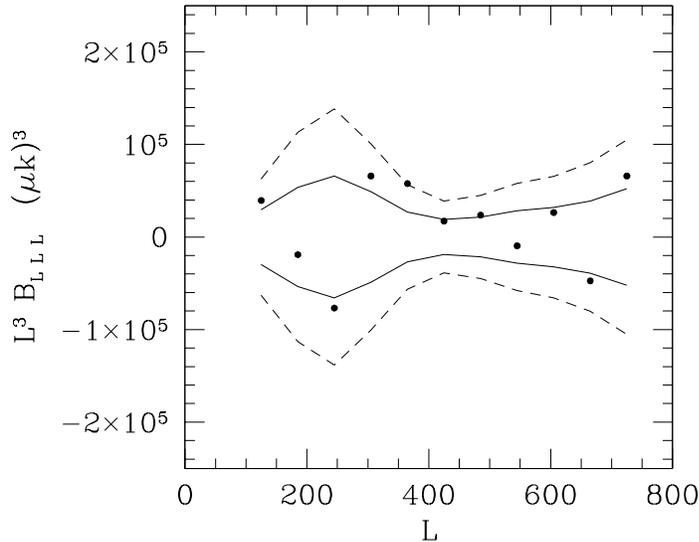}
\vspace{-1.5cm}
\end{center}
\caption{The minimum variance estimator - comparison with a Gaussian sky. The solid (dashed) line delimit
the $68\%$ ($95\%$) confidence region determined from a Monte Carlo
simulation of a Gaussian sky and the dots are the values obtained from the MAXIMA dataset.}
\label{fig_bspdiag}
\end{figure}

\begin{figure}
\begin{center}
\includegraphics[width=0.6\textwidth]{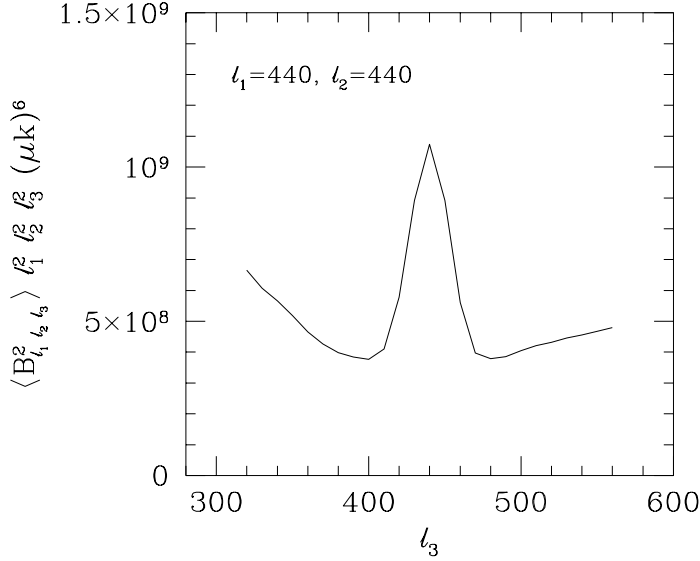}
\vspace{-1.5cm}
\end{center}
\caption{The variance of the bispectrum from the minimum variance estimator. Note the sharp
increase at $\ell_3=440$ due to the diagonal component.}
\label{fig_bspvar}
\end{figure}

The values obtained turn out to be too noisy and highly correlated, so that we
decided to combine them into larger bins of size $\Delta\ell=60$. We grouped
the values in the following way:
\bea
\label{mlbin}
\hat{B}_{L_1 L_2 L_3}={1\over N_{L_1 L_2 L_3}}\sum_{\ell_i\in I_i}
\hat{B}(\ell_1,\ell_2,\ell_3),
\eea
where  $I_i=[L_i-\Delta\ell/2,L_i+\Delta\ell/2]$ and $L_1\ge L_2\ge L_3$. 
$N_{L_1 L_2 L_3}$ is
the number of estimated bispectra that fall in the bin and $L_i$ will cover all
the multipole range with step $\Delta\ell$. Note that values obtained 
through this binning do not have
to necessarily obey the triangle relations for $L_i$.
We have a total of 216 binned values and applied the same process to the MC
maps to calculate the covariance matrix. In Fig. \ref{fig_bspdiag} we plot
the corresponding diagonal values together with the $68\%$ and $95\%$
confidence limits obtained if the sky was indeed Gaussian.
Note that this is only a small sample of all the bispectrum values.
The variance is expected to be smaller for the non-diagonal terms.
Although the variance for the estimator can only be obtained numerically, it
is possible to have a rough idea of why the $68\%$ contour follows the shape
shown in Fig. \ref{fig_bspdiag}. If non-Gaussianity is small, the cosmic variance
of the bispectrum, in the full sky setup, is given in terms of the two-point function of the 
$a_{\ell m}$ (\citealt{Luo94}; \citealt{Hea98}). The variance of the angle-averaged
bispectrum (eq. \ref{angle_average}) is then calculated as \citep{GM00})
\be
  \left<B_{\ell_1\ell_2\ell_3}^2\right> \approx
  C_{\ell_1}C_{\ell_2}C_{\ell_3}\Delta_{\ell_1\ell_2\ell_3},
\ee
where $\Delta_{\ell_1\ell_2\ell_3}$ takes the values 1, 2, and 6 
depending whether all $\ell$'s are different,
two of them are same or all are the same, respectively. 
$C_\ell$ is the total CMB angular power spectrum, including the noise.
On the other hand, the estimator we are using in the flat-sky approximation should correspond
approximately to an integration over rotations 
of products of three Fourier modes (see eq. \ref{bspdef}). We therefore expect 
its variance to follow the same features and be
approximately proportional to 
\be
\frac{1}{(\ell_1\ell_2\ell_3)^2}(\ell_1^2C_{\ell_1})
(\ell_2^2C_{\ell_2})(\ell_3^2C_{\ell_3})\Delta_{\ell_1\ell_2\ell_3}.
\ee
The quantity $\ell^2C_{\ell}$ is similar to the standard expression for the temperature
fluctuation for large multipoles: $\ell(\ell+1) C_\ell / 2\pi$, giving the Sachs-Wolfe
plateau. If we then plot $\ell_1^2\ell_2^2\ell_3^2\left<\hat{B}_{\ell_1\ell_2\ell_3}^2\right>$,
one should find a behavior close to the observed for the CMB power spectrum, multiplied
by the relevant $\Delta_{\ell_1\ell_2\ell_3}$. Figure \ref{fig_bspvar} shows exactly
that for $\ell_1=\ell_2=440$. One might think the peak at $\ell_3=440$ is
due to some sharp increase of the power spectrum, however an inspection of the 
current measured CMB power spectrum, shows this is not the case. Instead, the
increase is simply due to the fact all $\ell$'s are the same for the bispectrum,
so that $\Delta_{\ell_1\ell_2\ell_3}=6$, contrary to $\Delta_{\ell_1\ell_2\ell_3}=2$,
making the variance for the diagonal bispectrum three times bigger than the other values.
Also, if we look at fig. \ref{fig_bspdiag}, the $68\%$ contour should
be proportional to $(\ell^2 C_\ell)^{3/2}$ and follow the features of the
MAXIMA power spectrum (\citealt{hanany}): there is a peak at $\ell\approx 220$ and
a steady increase of the variance for large $\ell$ due to the increase of
the noise power spectrum.

\begin{figure}
\begin{center}
\includegraphics[width=1.0\textwidth]{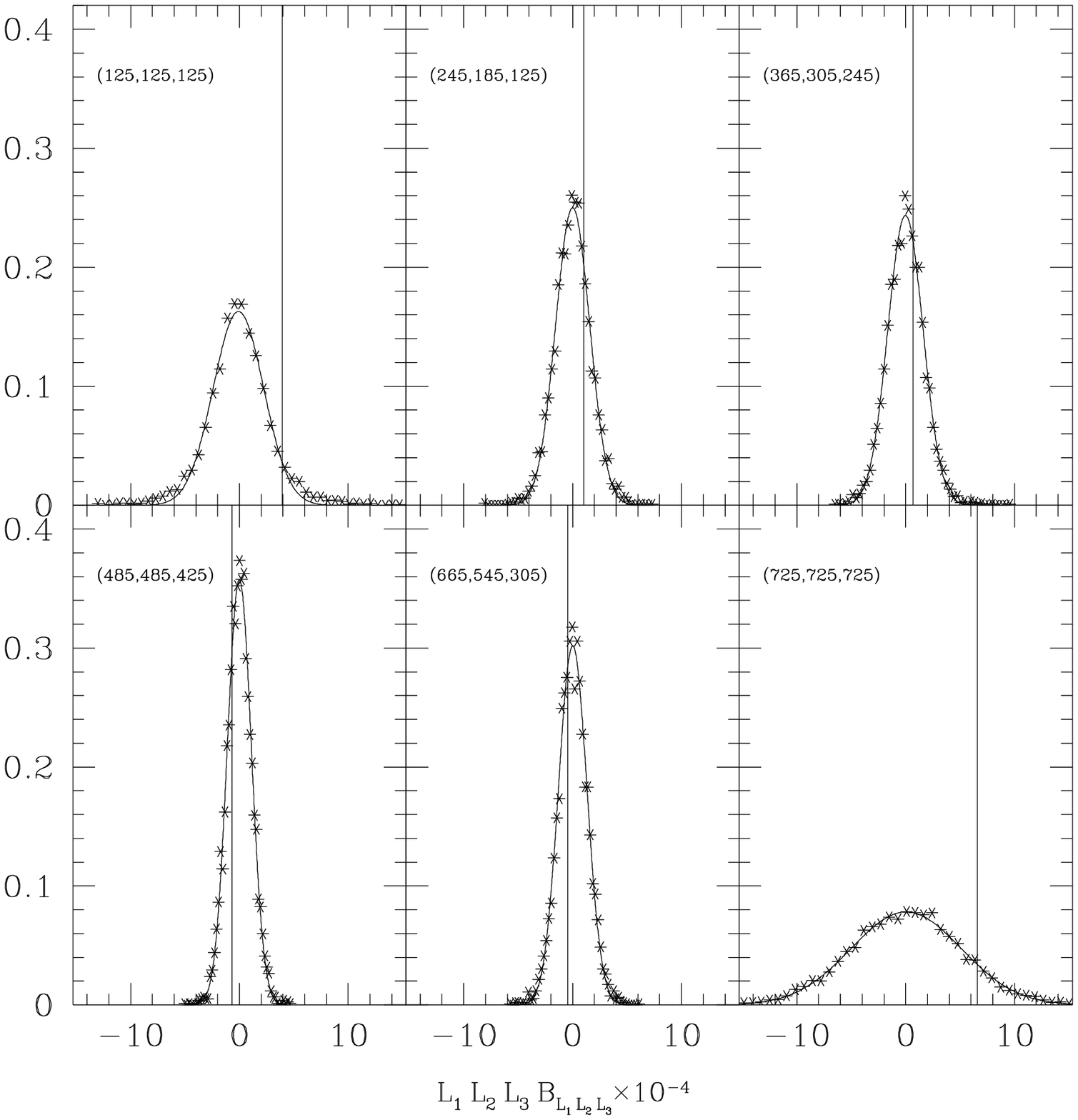}
\end{center}
\caption{Histograms of the minimum variance bispectrum estimator. Solid lines correspond
to the best fit Gaussian distribution. Vertical lines represent the
values obtained from the MAXIMA map.}
\label{fig_bsphist}
\end{figure}

\begin{figure}
\begin{center}
\includegraphics[width=0.6\textwidth]{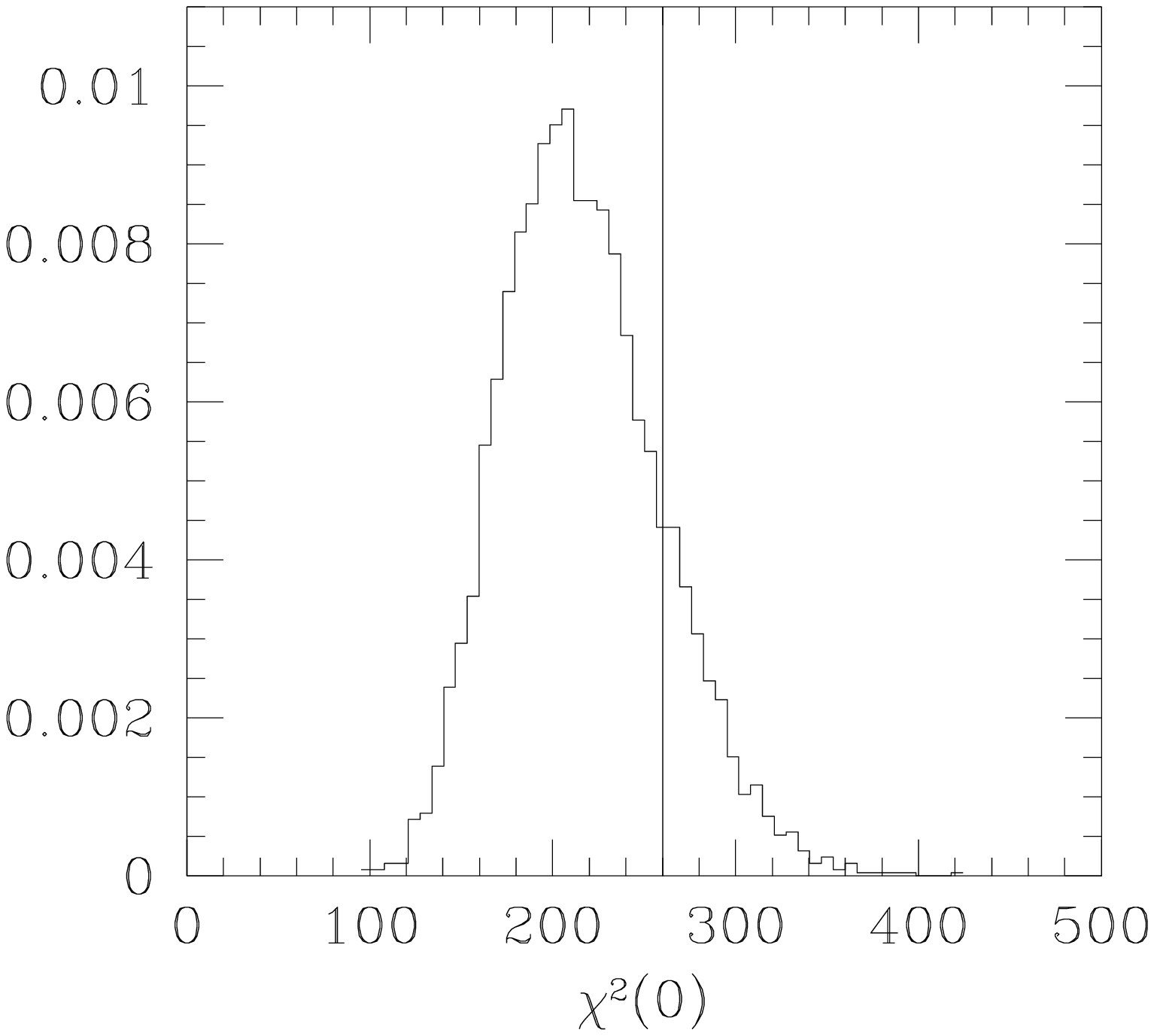}
\vspace{-1.5cm}
\end{center}
\caption{$\chi^2$ distribution (as defined in eq. \ref{chi2})
for the minimum variance estimator. The vertical line corresponds
to the measured $\chi^2$ (260).}
\label{fig_chi2}
\end{figure}

We would like to stress that there are two objectives in measuring
the bispectra. First we want to compare the estimates against the Gaussian
hypothesis and second we want to use this same measurements to constrain
theories of structure formation that generate non-Gaussianity.
We start by testing the Gaussian assumption.
If the temperature anisotropies have a Gaussian distribution then the
bispectrum estimates will obey
\be
\left<\hat{B}_{L_1 L_2 L_3}\right>=0
\ee
The values we actually measure should fluctuate around this average
in a way consistent with the probability distribution for the estimator
in a Gaussian sky. 
One way to test if all the values are indeed compatible
with the Gaussian hypothesis is through the use of the standard
$\chi^2$. This goodness of fit will be appropriate as long as
the distributions for the estimator are approximately Gaussian.
In Fig. \ref{fig_bsphist} we show some of the histogram distributions
for $\hat{B}_{L_1 L_2 L_3}$, obtained from the Monte-Carlo
realizations of the Gaussian sky. 
The first thing to note is that
indeed we have $\left<\hat{B}_{L_1 L_2 L_3}\right>=0$ as it should 
be for Gaussian temperature fluctuations, so that the estimator
is unbiased as expected. Most of the histograms are consistent
with a Gaussian distribution, although there are some deviations
for low multipoles, due to the low number of independent modes
contributing for these values. However these are only a small
number\footnote{4 out of 216 have ${95\%/68\%} > 2.2$, where
$95\%$ and $68\%$ correspond to the calculated confidence limits.}
and it should not spoil the use of the $\chi^2$
as a goodness of fit (an analysis without these values produced 
the same conclusions). Moreover, the measured $\chi^2$ can still
be a robust measure of non-Gaussianity, as long as we calculate its
full distribution from the MC realizations. We have thus considered the
quantity:
\be
\label{chi2}
\chi^2=\sum_{\alpha\alpha'}(\hat{B}^{obs}_{\alpha}
- B^{th}_{\alpha})\,C^{-1}_{\alpha \alpha'}\,
(\hat{B}^{obs}_{\alpha'}- B^{th}_{\alpha'}),
\ee
where $\alpha=(L_1,L_2,L_3)$, $B^{th}_{\alpha}=0$ and 
$C_{\alpha \alpha'}$ is the covariance matrix
of the estimators evaluated from the MC realizations.
From a total of 216 values we found $\chi^2_{obs}\approx 260$. Using
4000 MC maps we constructed the corresponding $\chi^2$ distribution
(Fig. \ref{fig_chi2}) and found the probability of the MC $\chi^2$
being larger than $\chi^2_{obs}$ to be ${\rm Prob}\,(\chi^2_{MC}>
\chi^2_{obs})\approx 15\%$.
Thus, the MAXIMA map seems to be reasonably consistent with the
Gaussian hypothesis.

Although there was no detection of clear deviation from Gaussianity,
one can still use the measured bispectrum to constrain theories
of structure formation. This subject will be specifically
addressed in a later section.

\section{The standard frequentist estimator}
\label{freq}

In the previous section we discussed an estimator for the bispectrum,
cubic in the temperature pixels, obtained by minimizing the
variance. This method deals automatically with partial sky coverage
and should have the smallest error bars. However, its numerical
implementation is somewhat complex and the computer calculations can
be quite slow. We can consider instead a more direct approach to
measure the bispectrum: do a harmonic transform or a Fourier transform
of the map and consider cubic products of the $a_{\ell m}$ or the
$a(\bk)$ as in equation (\ref{Blm}) or (\ref{bspdef}), combining them
in an appropriate way. Although this estimator will not necessarily
have the lowest variance possible, its numerical implementation is
quite straightforward and extremely fast, making the method more
feasible when dealing with very large datasets. We will therefore
review the so-called frequentist estimator, as introduced by
\citet{FMG98}. 
Although its basic description was already addressed in
\citet{santos}, it will be interesting to apply this method to exactly
the same map used in the previous section and see how well it will do
compared to the previous estimator.

In a full sky setup it is common to combine the measured $a_{\ell m}$ into the
angle-averaged bispectrum (see \citealt{KWS02} for a description of the method).
The relation to the reduced bispectrum was specified in equation (\ref{angle_average}). 
However, as already explained, it is more natural to use
the flat-sky approximation for the MAXIMA dataset.

\subsection{Flat case - FFT estimator}
\label{FFT}

The Fourier transform of the 3-point correlation function for the temperature
anisotropies measured by MAXIMA will not be given directly by equation
(\ref{bspdef}), but instead (see eq. \ref{fmode})
\bea
\label{bspexp}
\lefteqn{\langle a_s(\bk_1)a_s(\bk_2)a_s(\bk_3)\rangle=
\int\frac{d^2k'_1}{(2\pi)^2}\frac{d^2k'_2}{(2\pi)^2}\,\tilde{W}(\bk_1-\bk'_1)
\,\tilde{W}(\bk_2-\bk'_2)\,\tilde{W}(\bk_3+\bk'_1+\bk'_2)}
\nn & &
\times\left[\tilde{\cal B}(k'_1)\,\tilde{\cal B}(k'_2)\,
\tilde{\cal B}(|\bk'_1-\bk'_2|)\,
B(k'_1,k'_2,|\bk'_1-\bk'_2|)\right].
\eea
where it was assumed that the noise bispectrum is zero. The quantity in 
parenthesis should not change much along the integration, given the shape 
of the window function, and one can then simplify the integral:
\be
\label{bspexp2}
\langle a_s(\bk_1)a_s(\bk_2)a_s(\bk_3)\rangle\approx
\left[\tilde{\cal B}(k_1)\,\tilde{\cal B}(k_2)\,\tilde{\cal B}(k_3)\,
B(k_1,k_2,k_3)\right]\,{\cal V}\,,
\ee
with
\be
{\cal V}=\int {\rm d}^2x W^3(\bx)
\ee
and $\bk_1+\bk_2+\bk_3=0$.
So, to estimate the bispectrum, we should consider quantities like
\be
\label{estimate}
a_c(\bk_1)a_c(\bk_2)a_c(\bk_3)/{\cal V},
\ee
where $a_c(\bk)=a_s(\bk)/\tilde{\cal B}(k)$,
so that the corrections due to the finite size
of the map and the effects of the beam and pixelisation are already
taken into account. Furthermore, the bispectrum is invariant under
rotations and reflections, so that we need to combine values
contributing to the same bispectrum. To measure $B(\ell_1,\ell_2,
\ell_3)$, we could then use
\bea
\lefteqn{\hat{B}(\ell_1,\ell_2,\ell_3)={1\over {2 \pi \cal V}}
\int_0^{2\pi} {\rm d}\theta_1\,[a_c(\bk_1(\theta_1))
a_c(\bk_2(\theta_1,\theta))a_c(\bk_3(\theta_1,\theta))}
\nn & &
+a_c(\bk_1(\theta_1))a_c(\bk_2(\theta_1,-\theta))a_c(\bk_3(\theta_1,-\theta))],
\eea
where $\bk_1(\theta_1)=\ell_1(\cos\theta_1,\sin\theta_1)$, 
$\bk_2(\theta_1,\theta)=\ell_2(\cos(\theta_1+\theta),\sin(\theta_1+\theta))$,
$\bk_3(\theta_1,\theta)=-\bk_1(\theta_1)-\bk_2(\theta_1,\theta)$ and 
$\theta$ is such that $|\bk_3|=\ell_3$.
For each set of $(\ell_1,\ell_2,\ell_3)$ one would have in principle
to calculate several $a_s(\bk)$ so as to perform the integral
above. However, it is more efficient to first obtain all the Fourier
modes by doing a Fast Fourier Transform (FFT) of the map. The modes 
$a_s(\bk)$ will then be discretised on a grid with step size
$\Delta k$ determined by the width of the field of observation.
This is fine, since Fourier modes separated by less than $\Delta k$
will be highly correlated, but also means the bispectrum values will
be automatically discretised according to the grid.
Moreover, we still need to combine the estimates (eq. \ref{estimate})
related by rotations into the same measurement and to rearrange
the values into bins, so as to lower the noise and reduce the 
correlations. We therefore use the following bispectrum estimator:
\be
\label{freq_est}
\hat{B}_{L_1 L_2 L_3}={1\over N_{L_1 L_2 L_3}{\cal V}}
\sum_{|\bk_i|\in I_i}{\cal R}e\{a_c(\bk_1)
a_c(\bk_2)a_c(\bk_3)\},
\ee
where $I_i=[L_i-\Delta\ell/2,L_i+\Delta\ell/2]$, $N_{L_1 L_2 L_3}$ is
the number of estimated bispectra that fall in the bin and 
$\bk_1+\bk_2+\bk_3=0$.
$\Delta\ell$ is the size of the bin and again, the $L_i$ do not have
to obey the bispectrum selection rules.

\subsubsection{The data}

We applied the above estimator to exactly the same map used for the
best cubic estimator. Note that this map is not exactly rectangular
so that to use the Fast Fourier Transform we had to pad the map
with extra zeros. We took the extra bias introduced by this into
account when calculating the bispectrum.
The FFT assumes the data has periodic boundary conditions.
We therefore multiplied the map by a Welch window function
to correct for the mismatch at the border of the region.
This should reduce the leakage between neighboring Fourier modes
and from small to large scales. The use of the Welch window function
turns out to be essential to decrease the size of the error bars.
To obtain the probability distribution and covariance matrix of
the frequentist estimator in equation (\ref{freq_est}), we again
used Monte-Carlo simulations of the MAXIMA-1 dataset.

\subsubsection{Results}

\begin{figure}
\begin{center}
\includegraphics[width=0.6\textwidth]{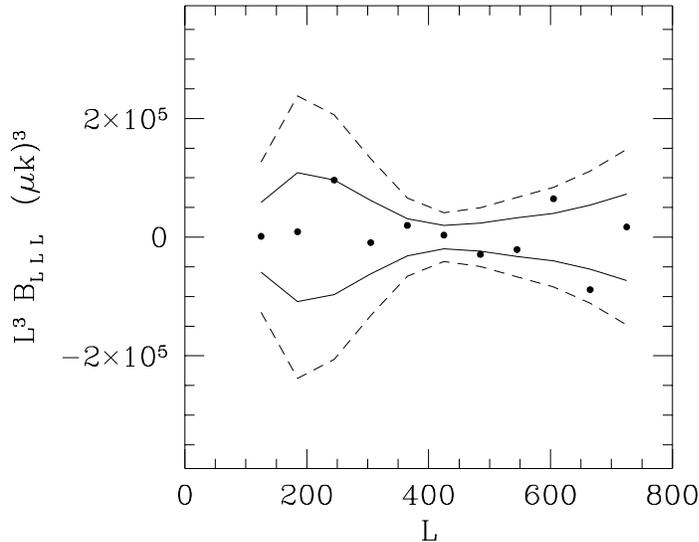}
\vspace{-1.5cm}
\end{center}
\caption{FFT estimator - diagonal bispectrum from the MAXIMA dataset compared to a Gaussian sky.}
\label{fig_bspdiag2}
\end{figure}

We applied the estimator in equation (\ref{freq_est}) to the data
using a bin size of $\Delta\ell=60$ from $L=125$ to $L=725$, so as to
match the same multipole values used for the binned best cubic
estimator (eq. \ref{mlbin}).
This gives a total
of 236 bispectrum values. We obtained the corresponding covariance
matrix by applying the same process to 10000 MC realizations.
In Fig. \ref{fig_bspdiag2} we plot
the corresponding diagonal values together with the 
$68\%$ and $95\%$ confidence limits.
Looking at the figure it seems that indeed the previous estimator
manages to produce smaller error bars than the FFT estimator.
This improvement on the signal to noise will also be confirmed
when looking at the constraints imposed by both estimators
to the weak non-linear coupling model.
Note that the variances of the diagonal components of the FFT estimator
presented here are slightly larger than in the previous analysis of \citet{santos}: 
we have found that this is primarily due to the fact that
we are considering smaller bins and different regions of the map. 

\begin{figure}
\begin{center}
\includegraphics[width=0.6\textwidth]{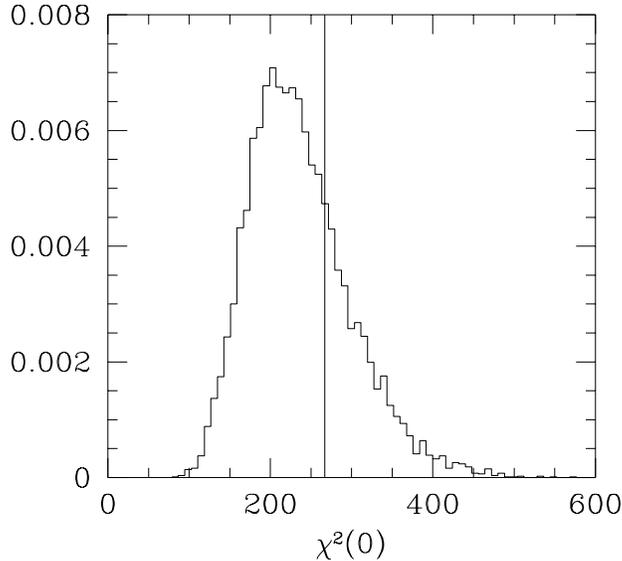}
\vspace{-1.5cm}
\end{center}
\caption{$\chi^2$ distribution for the FFT estimator. Vertical line
corresponds to $\chi^2\sim 267$.}
\label{fig_chi2b}
\end{figure}

To test the Gaussian assumption we again considered the standard
$\chi^2$ as defined in equation (\ref{chi2}). From the MC realizations
we checked that the estimator is indeed unbiased and has an 
histogram distribution approximately Gaussian.
From a total of 236 values we found $\chi^2_{obs}\approx 267$. Using
10000 MC maps we constructed the corresponding $\chi^2$ distribution
(Fig. \ref{fig_chi2b}) and found ${\rm Prob}\,(\chi^2_{MC}>
\chi^2_{obs})\approx 27\%$. 
The frequentist estimator also gives results consistent with the Gaussian
assumption.

\section{Cosmological Implications}
\label{cosmo}

We have been discussing the use of the bispectrum estimates as a way to test
for Gaussianity of the MAXIMA-1 dataset. However the bispectrum can be
applied in more powerful ways: several theories of structure formation
give specific, clean predictions for the bispectrum, which can then be
compared to the data in much the same way as it is done for the power
spectrum. Even if the temperature anisotropies turn out to be consistent with a Gaussian
distribution, we can always use the measurements to put constraints
on the parameters of these theories.

Although we are interested in the bispectrum from a primordial origin,
the signal can be contaminated with various foregrounds, such as
extra-galactic radio and infrared sources, the Sunyaev-Zel'dovich (SZ) effect,
the weak lensing effect and so on. For a discussion of these sources
see \citet{CH00}. The Ostriker-Vishniac effect and the coupling between 
the SZ and the weak lensing effects, typically give rise to non-zero
bispectra on very small angular scales.
Therefore they should not be detectable by MAXIMA, with the map
we have been considering.
The patch of the sky analyzed
in the MAXIMA-1 experiment \citep{hanany}, shows no detectable
radio or infrared sources. Moreover, these point sources would
produce a reduced bispectrum approximately constant \citep{KS01},
which would show up clearly in Fig. \ref{fig_bspdiag} (or
Fig. \ref{fig_bspdiag2}) as a sharp rise at high $\ell$. Finally,
the contributions from interstellar dust and synchrotron radiation
were shown to be negligible.
From now on we will assume there are no measurable foreground
contributions to the bispectrum and concentrate on the possible
primordial sources , which can give rise to non-Gaussianity on
degree and sub-degree scales. Even if there were such contributions, a 
cancellation of the several signals is unlikely to occur and the
constraints obtained should then give upper limits.

The main theory for structure formation in the universe is Inflation 
and, as already stated, it predicts Gaussian fluctuations to a very
good approximation. This can be tracked down to the requirement
that the inflaton potential must be flat enough so that Inflation
can occur. The perturbations in the inflaton field, which are 
responsible for the generation of cosmological fluctuations, should
then be linear. This basically means we are neglecting any interaction
between the perturbed inflaton and other fields (or itself).
However, if we took these interactions into account, the different
Fourier modes of the above vacuum perturbation would couple, 
leading to non-Gaussian cosmological fluctuations.
A few authors have considered this possibility and calculated
the 3-point correlation function of the inflaton field (\citealt{GLMM94};
\citealt{FRS93}),
which in turn generates a non-zero bispectrum. Typically the 
bispectrum obtained in these calculations, can be directly 
created just by considering a simple weak non-linear coupling
model for the primordial curvature perturbation \citep{KS01}:
\be
  \label{curv_model}
  \Phi({\mathbf x})
 =\Phi_L({\mathbf x})
 +f_{NL}\left(
              \Phi^2_{L}({\mathbf x})-
	      \left<\Phi^2_{L}({\mathbf x})\right>
        \right),
\ee
where $\Phi({\bf x})$ is the curvature perturbation in real
space and $\Phi_L({\bf x})$ the Gaussian part of the perturbation
($\left<\Phi_L({\bf x})\right>=0$). The non-linear coupling constant,
$f_{NL}$, will be related to the slope and curvature of the inflaton
potential \citep{SB90,GLMM94}. For example, $f_{NL}\approx 3\epsilon
-2\eta$, where $\epsilon$ and $\eta$ are the slow-roll parameters and
typically  $f_{NL}\sim 10^{-2}$. The dominant source of
non-Gaussianity is the weak non-Gaussian gravitational growth of
perturbations at early times: general relativistic
second-order perturbation theory gives $f_{NL}\sim {\cal O}(1)$
\citep{PC96}.
More drastic changes to the evolution of the inflaton field,
as long as they are brief, could
still be compatible with Inflation and generate strong 
non-Gaussianity.  An example of a potentially strongly non-Gaussian
model is the curvaton model of \citet{LW2002}.  
Some recent proposals for using a primordial
power spectrum with a break or a bump \citep{B01,GSZ00}
can give a good fit to the current CMB data
and the non-Gaussian features of such models could be used
as a way to break the degeneracy with standard cosmological models. 
\citet{WK00} analyzed
a simple model for an inflaton potential with a feature
and showed the level of non-Gaussianity is still negligible,
but it would be interesting to further explore this possibility.
Contrary to standard adiabatic perturbations, isocurvature ones
can easily generate non-Gaussianity. They are normally created
during Inflation by vacuum fluctuations of a non-inflaton
field, so that the flatness requirements for the potential no
longer apply. Pure isocurvature perturbations are however ruled
out by experiment \citep{trd01}. \citet{BMR02} considered instead
correlated adiabatic and entropy perturbations, which should
lead to sizeable non-Gaussian features in both modes.
Topological defects are also well known to generate non-Gaussianity.
Cosmic strings, for instance, can produce non-Gaussian temperature
fluctuations through weak lensing: the Kaiser-Stebbins effect \citep{ks84}.
A proposal to promote the scale invariance of the power 
spectrum to full conformal invariance \citep{AMM97} also
accommodates interesting non-Gaussian possibilities.
In the current analysis we shall concentrate on the model described
by equation (\ref{curv_model}) and only consider terms linear in $f_{NL}$.

\subsection{Weak non-linear coupling}

In the case of primordial adiabatic scalar perturbations, the 
coefficients $a_{\ell m}$ can be written as \citep{MB95}:
\be
  \label{alm_curv}
  a_{\ell m}=4\pi(-i)^\ell
  \int\frac{d^3k}{(2\pi)^3}\Phi({\bf k})\Delta_{\ell}(k)
  Y_{\ell m}^*(\hat{\bf k}),
\ee
where $\Delta_{\ell}(k)$ is the radiation transfer function. 
One could also calculate the isocurvature perturbation by 
considering instead the primordial entropy perturbation,
the corresponding transfer function and possibly, an
expression for this primordial fluctuation similar to
equation (\ref{curv_model}).
From this,
one can then calculate the reduced bispectrum \citep{KS01}
\bea
  b_{\ell_1\ell_2\ell_3}^{theory}
  \label{blll_theory}
  = 2 f_{NL}\int_0^\infty r^2 dr 
    \left[
          b^L_{\ell_1}(r)b^L_{\ell_2}(r)b^{NL}_{\ell_3}(r)+
	  b^L_{\ell_1}(r)b^{NL}_{\ell_2}(r)b^{L}_{\ell_3}(r)+
	  b^{NL}_{\ell_1}(r)b^L_{\ell_2}(r)b^{L}_{\ell_3}(r)
    \right],
\eea
where 
\bea
  \label{bL}
  b^L_{\ell}(r) &\equiv&
  \frac2{\pi}\int_0^\infty k^2 dk P_\Phi(k)\Delta_{\ell}(k)j_\ell(kr),\\
  \label{bNL}
  b^{NL}_{\ell}(r) &\equiv&
  \frac2{\pi}\int_0^\infty k^2 dk \Delta_{\ell}(k)j_\ell(kr).
\eea
$P_\Phi(k)$ is the primordial matter power spectrum and 
$b_{\ell_1\ell_2\ell_3}^{theory}$ will have only one free parameter - $f_{NL}$.
All other cosmological parameters will be fixed by fitting the CMB power 
spectrum, $C_\ell$ to the data.
We used CMBFAST \citep{cmbfast} to obtain the radiation transfer function and
calculate the integrals in equation (\ref{bL}) and (\ref{bNL})\footnote{The 
non existence of a $P_\Phi(k)$ term in equation (\ref{bNL}) requires some
care when obtaining the transfer function: one should increase the limits
of the time integration used to calculate $\Delta_{\ell}(k)$.}.
We generated the theoretical bispectrum by running this modified version of CMBFAST
with the best fit parameters obtained
in \citet{balbi} and normalised $P_\Phi(k)$ to the MAXIMA-1 and COBE/DMR data.
Note that $b_{\ell_1\ell_2\ell_3}^{theory}$ (eq. \ref{blll_theory}) will
depend on the square of the normalization used for the power spectrum, so
that the choice of normalization will affect the value obtained for 
$f_{NL}$.
Also, when generating these values, we consider $f_{NL}=1$.
\begin{figure}
\begin{center}
\includegraphics[width=0.6\textwidth,height=0.6\textwidth]{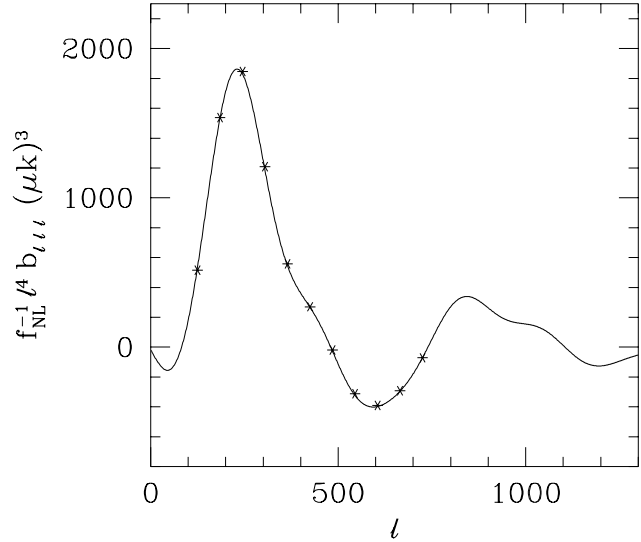}
\end{center}
\vspace{-1.5cm}
\caption{Diagonal theoretical bispectrum. The stars indicate the theoretical
values used to compare to experiment, after the relevant binning (eq. \ref{mlbin}).}
\label{fig_diagth}
\end{figure} 
In Fig. \ref{fig_diagth} we show the diagonal components for the theoretical
reduced bispectrum. We multiplied the values by $\ell^4$, which basically
corresponds to introducing a factor of $\ell(\ell+1)$ for each $b^L_{\ell}(r)$,
making it approximately constant for low $\ell$ as in the case of the $C_\ell$.
Note that this is no longer possible for the non-diagonal components, due to 
the mixed multipoles (eq. \ref{blll_theory}) and there is no standard factor
we could use in that case. 
We are now in conditions to compare the theoretical values with the ones
estimated from the data.

\subsubsection{Minimum variance estimator}

\begin{figure}
\begin{center}
\includegraphics[width=0.6\textwidth,height=0.6\textwidth]{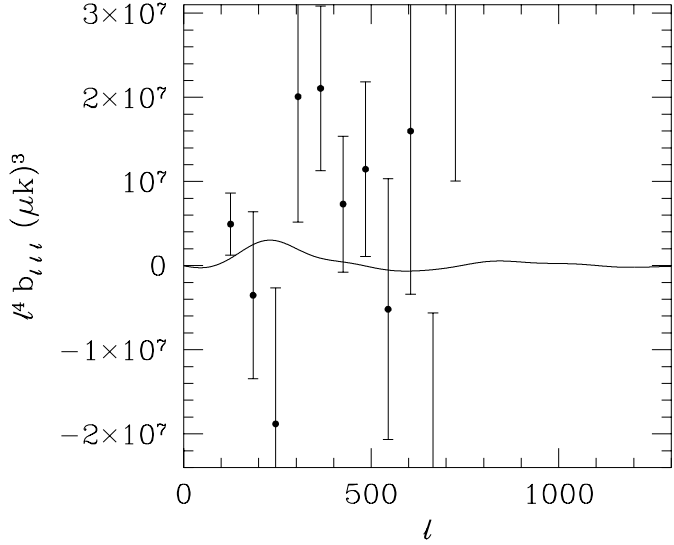}
\end{center}
\vspace{-1.5cm}
\caption{Comparison between theory and experiment. The data points were obtained using the
minimum variance estimator.}
\label{fig_thexp}
\end{figure}

\begin{figure}
\begin{center}
\includegraphics[width=0.6\textwidth,height=0.55\textwidth]{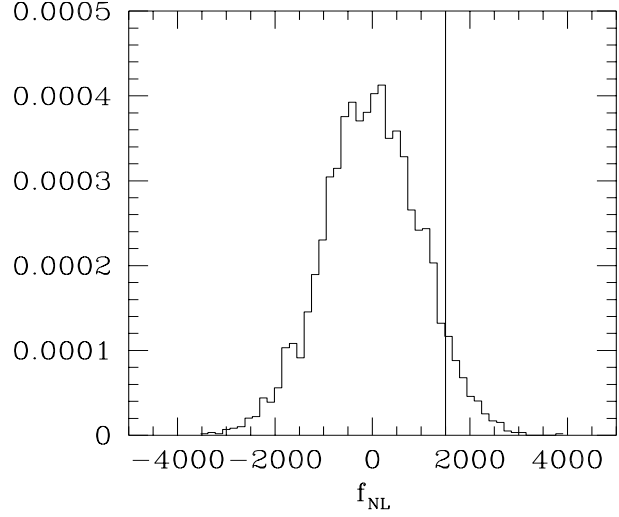}
\end{center}
\vspace{-1.5cm}
\caption{$f_{NL}$ distribution for the minimum variance estimator.}
\label{fig_fdist}
\end{figure}

We should in principle use equation (\ref{ytobsp_norm}) for each value
to translate from the theoretical bispectrum to the estimated
bispectrum. Unfortunately this is not practical since it would require
the full knowledge of the window function, 
$w(\ell_1,\ell_2,\ell_3,\ell_1',\ell_2',\ell_3')$. However, as can
be seen in Fig. \ref{fig_diagth}, the
theoretical bispectrum changes quite slowly along the width of the
window function (see fig. \ref{fig_window}), so that
we can safely extract the bispectrum from the integral (which then 
gives just 1).
So, basically, we generated the theoretical bispectrum for each of the 
modes analyzed in section \ref{mlresults} (1409 in total) and combined
them into bins in exactly the same way as before, using equation
(\ref{mlbin}). We then fit the free parameter, $f_{NL}$ to the data
by using the standard $\chi^2$ statistics:
\be
\label{chi2_fnl}
\chi^2(f_{NL})=\sum_{\alpha,\alpha'}(\hat{B}^{obs}_{\alpha}
- f_{NL}\,B^{th}_{\alpha})\,C^{-1}_{\alpha \alpha'}\,
(\hat{B}^{obs}_{\alpha'}- f_{NL}\,B^{th}_{\alpha'}),
\ee
where again, $C_{\alpha \alpha'}$ is the covariance matrix of the
estimators obtained from the Monte-Carlo realizations.
$B^{th}_{\alpha}$ is the binned, calculated bispectrum from equation
(\ref{blll_theory}) without the $f_{NL}$.
Assuming a Gaussian distribution for the estimated bispectrum, we
can obtain the maximum probability for $f_{NL}$ just by minimizing
the above $\chi^2$ with respect to $f_{NL}$:
\be
f_{NL}=\left[\sum_{\alpha,\alpha'}\hat{B}^{obs}_{\alpha}
C^{-1}_{\alpha \alpha'}B^{th}_{\alpha'}\right]\,
\left[\sum_{\alpha,\alpha'}\hat{B}^{th}_{\alpha}
C^{-1}_{\alpha \alpha'}B^{th}_{\alpha'}\right]^{-1}
\ee
The best fit value is $f_{NL}\sim 1500$. 
Using this value, we show in Fig. \ref{fig_thexp} an 
interesting comparison between theory and experiment.
It is clear that the estimated error is still too large to allow
us to make a clear statement about the value of $f_{NL}$.
Even $f_{NL}=0$ seems reasonable when looking at this plot.
However we expect the non-diagonal terms to have smaller
error bars, which may bias the data to prefer a larger value
for $f_{NL}$.
Applying the same
minimization to each Monte-Carlo realization we obtain
the distribution for the non-linear coupling parameter, $f_{NL}$.
This is shown in Fig. \ref{fig_fdist}.
The peak of the distribution is at zero, as expected for 
MC simulations of a Gaussian sky, so that the estimated
$f_{NL}$ obtained by minimization of the standard $\chi^2$
is in fact unbiased. The $68\%$ confidence limit gives 
$|f_{NL}|\lsim 950$. Although the analysis seems to indicate a
value for $f_{NL}$ slightly larger than what one would expect
for a Gaussian map, the discrepancy is not strong enough for
a claim of detection of a non-zero $f_{NL}$ (approximately
$20\%$ of the values would be larger than the one obtained).

\subsubsection{FFT estimator}

\begin{figure}
\begin{center}
\includegraphics[width=0.6\textwidth,height=0.55\textwidth]{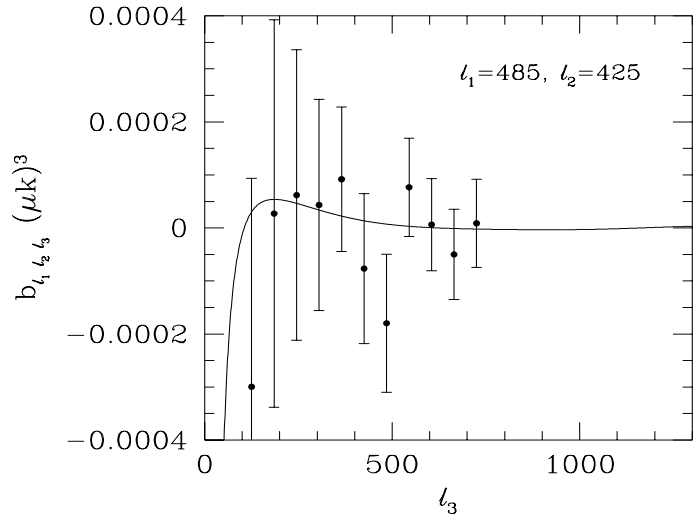}
\end{center}
\vspace{-1.5cm}
\caption{Comparison between theory and experiment for the FFT
estimator (non-diagonal components).
The combination used for $\ell_1$ and $\ell_2$ is supposed to give the
smallest error bars.}
\label{fig_thexp2}
\end{figure}

\begin{figure}
\begin{center}
\includegraphics[width=0.6\textwidth]{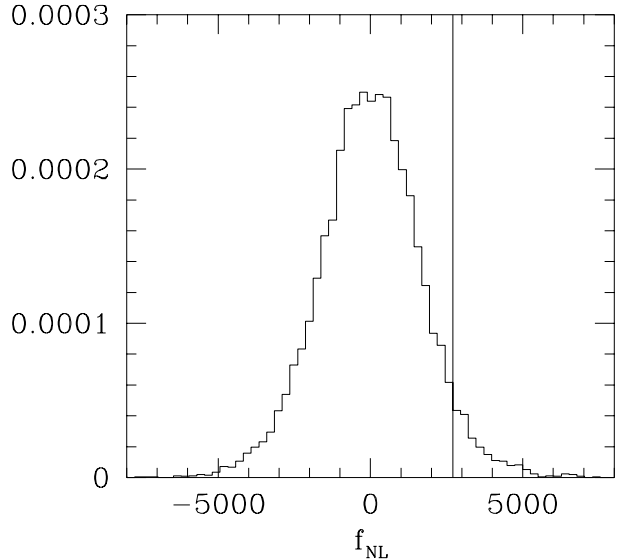}
\end{center}
\vspace{-1.5cm}
\caption{$f_{NL}$ distribution for the FFT estimator.}
\label{fig_fdist2}
\end{figure}

The analysis using the data obtained by the frequentist estimator
is quite similar to the one above. We again generate the
theoretical bispectrum for each of the modes measured by this
estimator and combine the values into bins in exactly the same
way as in section \ref{FFT}.
The best fit value gives $f_{NL}\sim 2700$. 
Using these fit, we plot in Fig. \ref{fig_thexp2} the experimental values
against the non-diagonal components of the theoretical bispectrum.
The corresponding
histogram distribution for $f_{NL}$ is shown in Fig. \ref{fig_fdist2}.
The $68\%$ confidence limit gives $|f_{NL}|\lsim 1650$.
As expected the constraint on a non-zero $f_{NL}$ is larger
than in the minimum variance estimator.  This is a slightly weaker
limit than obtained by \citet{santos}, but the previous limit
assumed only a Sachs-Wolfe contribution to the bispectrum, whereas
this analysis includes the full radiation transfer function.

\section{Discussion and Conclusions}
\label{discussion}

In this paper we have developed a minimum variance estimator
for the flat sky approximation in the limit of weak non-Gaussianity. 
We have discussed its numerical implementation
and substantiated some approximations that make the method faster. As a test
we applied this estimator to the MAXIMA-1 CMB map and obtained a total of 
1409 bispectrum modes combined into 216 bins. An analysis of these values
using a $\chi^2$ statistics indicates that the data is consistent with
Gaussianity. We also reviewed the more simple FFT estimator and
applied it to the same dataset so as to compare the two methods.
Although the overall conclusions obtained from an analysis of the CMB
map are the same, the FFT estimator shows a larger variance 
and higher correlations between the bispectrum modes 
than the previous method. This should convince us to look at the 
minimum variance cubic estimator as the method of choice to measure the
bispectrum. However, contrary to the FFT estimator, this method
is quite slow in computational terms making it more difficult to apply
to large datasets. Fortunately the study presented in this paper also 
shows a possible way out: one should still use the standard frequentist method
(equation \ref{freq_est}) but apply it instead to the inverse map,
$z_i=\xi_{ij}^{-1}\Delta T({\bx_j})$, where the 2-point
correlation function of the temperature field, $\xi_{ij}$, is
already known from the power spectrum measurements. The obtained
values should then be compensated by the relevant factor, e.g. 
equation (\ref{M_use}), to obtain the bispectrum.
This makes the minimum variance estimator method viable to use with the large
datasets from the next generation satellite experiments, the Microwave
Anisotropy Probe (MAP) and Planck.

We also used the measured bispectra to constrain cosmological
models. We considered a primordial non-Gaussian curvature perturbation
and generated the corresponding bispectrum taking into account the
full radiation transfer function with the best fit cosmological parameters from the
MAXIMA-1 experiment. We obtained a weak  
constraint on the non-linear coupling parameter, $f_{NL}$. 
The 68\%
confidence limit from the minimum variance estimator is $|f_{NL}|<950$.
This is the tightest limit on primordial non-Gaussianity in the CMB
which includes the full effects of the radiation transfer function.
The fit to the
data from both methods, seems to indicate a value for $f_{NL}$ slightly 
larger than what one would expect from a Gaussian sky.
If we were to ascribe this to a cosmological origin it would mean
the slow-roll conditions during inflation were violated.
However, the value obtained could be due to some foreground 
contamination, or just due to a statistical
fluctuation (around $20\%$ of all Gaussian skies could still yield
a value for $|f_{NL}|$ larger than the one obtained).

The current variance for the bispectrum is still too large to allow a
clean measurement of the bispectrum. The situation should improve with
current and future satellite experiments and using the methods
developed in this paper one will hopefully be able to put stringent
constraints on $f_{NL}$.  One should bear in mind, however, that
neither the MAP or Planck experiments will be able to detect
primordial non-Gaussianity with the usual slow-roll assumption for Inflation.

\section*{ACKNOWLEDGEMENTS}

We would like to thank Eiichiro Komatsu, Jo\~ao Magueijo
and Lloyd Knox for useful discussions and comments.
MGS acknowledges support from
Funda\c{c}\~ao
para a Ci\^encia e a Tecnologia.
JHPW and AHJ acknowledge support from
NASA LTSA Grant no.\ NAG5-6552 and NSF KDI Grant no.\ 9872979.
PGF acknowledges support from the Royal Society.
RS and SH acknowledge support from
NASA Grant NAG5-3941.
BR and CDW acknowledge support from NASA GSRP
Grants no.\ S00-GSRP-032 and S00-GSRP-031.
MAXIMA is supported by NASA Grant
NAG5-4454 and by the NSF through the CfPA
at UC Berkeley, NSF cooperative agreement AST-9120005.

\appendix

\section{All sky and Flat sky Bispectrum}
\label{bsp_flat}

We want to establish the correspondence between the full sky bispectrum
and the flat sky bispectrum. The approximation will be valid
when analysing a small patch of the sky and so, to correctly
establish the relation, we should take into account the effect
of incomplete sky coverage on the bispectrum estimator.

\subsection{Incomplete sky coverage}
Incomplete sky coverage means the harmonic coefficients
will be given by,
\be
a_{\ell m}^s=\int d\Omega \Delta T(\Omega)W(\Omega)Y_{\ell m}^*(\Omega)=
\int_{\Omega_{obs}} d\Omega \Delta T(\Omega)Y_{\ell m}^*(\Omega),
\ee
where $W(\Omega)$ defines the region of observation (1 in the region
and 0 outside) and $\Omega_{obs}$ denotes the solid angle of the
observed sky. Orthonormality of the spherical harmonics on the sky
is therefore destroyed and the $a_{\ell m}^s$ will be related to the
true harmonic transform, $a_{\ell m}$, through
\be
a_{\ell m}^s=\sum_{\ell'=0}^\infty\sum_{m'=-\ell'}^{\ell'}
a_{\ell m}
\int_{\Omega_{obs}} d\Omega Y_{\ell' m'}(\Omega)Y_{\ell m}^*(\Omega).
\ee
The estimated bispectrum will then be biased compared to
the true bispectrum:
\bea
\langle a_{\ell_1 m_1}^s a_{\ell_2 m_2}^s a_{\ell_3 m_3}^s\rangle=
\sum_{all\ \ell'}b_{\ell'_1 \ell'_2 \ell'_3}\sum_{all\ m'}
\int {\rm d}\Omega\,Y_{\ell'_1 m'_1}^*(\Omega)Y_{\ell'_2 m'_2}^*(\Omega)
Y_{\ell'_3 m'_3}^*(\Omega) & & \nonumber \\
\times \int d\Omega_1 W(\Omega_1)Y_{\ell'_1 m'_1}(\Omega_1)
Y_{\ell_1 m_1}^*(\Omega_1) & & \nonumber \\
\times \int d\Omega_2 W(\Omega_2)Y_{\ell'_2 m'_2}(\Omega_2)
Y_{\ell_2 m_2}^*(\Omega_2) & & \nonumber \\
\times \int d\Omega_3 W(\Omega_3)Y_{\ell'_3 m'_3}(\Omega_3)
Y_{\ell_3 m_3}^*(\Omega_3) & & \nonumber \\
\approx b_{\ell_1 \ell_2 \ell_3} \int d\Omega W^3(\Omega)
\,Y_{\ell_1 m_1}^*(\Omega)Y_{\ell_2 m_2}^*(\Omega)
Y_{\ell_3 m_3}^*(\Omega).
\label{incomp_bsp}
\eea 
Where we have used,
\be
\langle a_{\ell_1 m_1} a_{\ell_2 m_2} a_{\ell_3 m_3}\rangle=
b_{\ell_1 \ell_2 \ell_3}
\int {\rm d}\Omega\,Y_{\ell_1 m_1}^*(\Omega)Y_{\ell_2 m_2}^*(\Omega)
Y_{\ell_3 m_3}^*(\Omega)
\ee
and $\sum_{\ell m} Y_{\ell
m}(\Omega)Y_{\ell m}^*(\Omega') = \delta(\Omega-\Omega')$.
One also needs to assume that $b_{\ell_1 \ell_2 \ell_3}$ varies 
much more slowly than
the coupling integral, so that we can take $b_{\ell_1 \ell_2 \ell_3}$
outside the sum over $\ell'$ (the sum over $m'$ should peak at
$\ell'_1=\ell_1$, $\ell'_2=\ell_2$, $\ell'_3=\ell_3$).
This is the expression we should be comparing to the flat case.
If we further convolve equation (\ref{incomp_bsp}) with the 
Wigner-3j symbol, 
\be
\left({\ell_1\ \ell_2\
\ell_3}\atop{m_1\ m_2\ m_3}\right),
\ee
it is possible to obtain the corresponding bias for the angular
averaged bispectrum (\citealt{KWS02}):
\be
\langle B^s_{\ell_1\ell_2\ell_3}\rangle\approx
\frac{\Omega_{obs}}{4\pi} B_{\ell_1\ell_2\ell_3},
\ee
so that the correction for the angular averaged bispectrum corresponds,
as expected,
approximately to the fraction of the sky observed.
For the flat sky analysis, the measured Fourier modes will
also be given by a convolution of the true ones, through
\be
a_s(\bk)=\int\frac{d^2k'}{(2\pi)^2}\,\tilde{W}(\bk-\bk')
a({\bk'}).
\ee
Using equation (\ref{bspdef}) we then have,
\bea
\lefteqn{\langle a_s(\bk_1)a_s(\bk_2)a_s(\bk_3)\rangle\approx
B({k_1},{k_2},{k_3})}
\nn & &\times
\int\frac{d^2k'_1}{(2\pi)^2}\frac{d^2k'_2}{(2\pi)^2}\,\tilde{W}(\bk_1-\bk'_1)
\,\tilde{W}(\bk_2-\bk'_2)\,\tilde{W}(\bk_3+\bk'_1+\bk'_2)
\nn & & =
B(k_1,k_2,k_3)\int d^2x\,W^3(\bx)e^{-i\bx\cdot(\bk_1+\bk_2+\bk_3)},
\label{incomp_fbsp}
\eea
where again it was assumed the window function is sharp enough so that
we can take $B(k_1,k_2,k_3)$ outside the integral.
We are now in conditions to prove the correspondence between
the all sky and the flat sky bispectrum.

\subsection{All sky to Flat sky correspondence}
As in \citet{H00}, let us start by showing
the relation between the Fourier modes, $a({\mathbf k})$ and the
harmonic coefficients, $a_{\ell m}$.
For small angles around the pole ($\theta \rightarrow 0$), one can 
approximate the spherical harmonics by
\be
Y_{\ell m}(\theta,\phi)\approx J_m(\ell \theta)\sqrt{\frac{\ell}{2\pi}}
e^{i m \phi},
\ee
where $J_m(\ell \theta)$ is a Bessel function. We can then relate the spherical
harmonics with the plane wave, using the expansion:
\bea
\label{delta2harm}
e^{i \bmath{\ell}\cdot\mathbf{x}}=\sum_m i^m J_m(\ell x) 
e^{i m(\phi_x-\phi_{\ell})} \approx & & \nonumber \\
\sqrt{\frac{2\pi}{\ell}}\sum_m{i^m Y_{\ell m}(x,\phi_x)
e^{-i m\phi_{\ell}}},
\eea
where $\bmath{\ell}=(\ell\cos\phi_{\ell},\ell\sin\phi_{\ell})$ and
$\ell$ can be seen as the continuous limit of the integer labeling
the multipoles $a_{\ell m}$. This is valid for small $x$ where
$\mathbf{x}=(x\cos(\phi_x),x\sin(\phi_x))$. Near the pole of 
spherical coordinates, defined by $(\theta,\phi)$, we can lay
down almost Cartesian coordinates by defining the 2-dimensional
vector $\bmath{\theta}=(\theta\cos(\phi),\theta\sin(\phi))$, and then
use the correspondence $\mathbf{x}=\bmath{\theta}$,
as already implied in the above expression.
The inverse relation is then,
\be
Y_{\ell m}(x,\phi_x)=\sqrt{\frac{\ell}{2\pi}}i^{-m}
\int_0^{2\pi}\frac{d\phi_{\ell}}{2\pi}e^{im\phi_{\ell}}
e^{i \bmath{\ell}\cdot\mathbf{x}}.
\ee
Thus,
\bea
\lefteqn{\Delta T(\Omega)=\sum_{\ell m}a_{\ell m}Y_{\ell m}(\theta,\phi)}
\nn & & \approx
\sum_{\ell m}a_{\ell m}\sqrt{\frac{\ell}{2\pi}}i^{-m}
\int_0^{2\pi}\frac{d\phi_{\ell}}{2\pi}e^{im\phi_{\ell}}
e^{i \bmath{\ell}\cdot\bmath{\theta}}
\nn & & \approx
\int{\frac{d^2\ell}{(2\pi)^2}\left[\sqrt{\frac{2\pi}{\ell}}\sum_m i^{-m}
a_{\ell m}e^{im\phi_{\ell}}\right]e^{i \bmath{\ell}\cdot\bmath{\theta}}}.
\eea
The corresponding Fourier mode should then be given by
\be
a(\bmath{\ell})=\sqrt{\frac{2\pi}{\ell}}\sum_m i^{-m}
a_{\ell m}e^{im\phi_{\ell}},
\ee
for large $\ell$.
Its inverse relation is:
\be
\label{alm2ak}
a_{\ell m}=\sqrt{\frac{\ell}{2\pi}}i^m\int\frac{d\phi_\ell}{2\pi}
e^{-im\phi_{\ell}}a(\bmath{\ell}).
\ee

Finally, we can now work out the connection between the all sky and
flat sky bispectrum. Using the above relation,
\bea
\lefteqn{\langle a_{\ell_1 m_1}^s a_{\ell_2 m_2}^s a_{\ell_3 m_3}^s\rangle\approx
\sqrt{\frac{\ell_1}{2\pi}}\sqrt{\frac{\ell_2}{2\pi}}
\sqrt{\frac{\ell_3}{2\pi}}i^{m_1+m_2+m_3} 
B(\ell_1,\ell_2,\ell_3)}
\nn & &
\times\int\frac{d\phi_{\ell_1}}{2\pi}\frac{d\phi_{\ell_2}}{2\pi}
\frac{d\phi_{\ell_3}}{2\pi}e^{-im_1\phi_{\ell_1}}e^{-im_2\phi_{\ell_2}}
e^{-im_3\phi_{\ell_3}}
\int d^2x\,W^3(\bx)e^{-i\bx\cdot(\bmath{\ell}_1+\bmath{\ell}_2+\bmath{\ell}_3)},
\eea
where equations (\ref{alm2ak}) and (\ref{incomp_fbsp}) were used.
The integration involving the exponential,
$e^{-i\bx\cdot(\bmath{\ell}_1+\bmath{\ell}_2+\bmath{\ell}_3)}$,
will be done only for small values of $x$, due to the presence of the 
window function, so that it is a good approximation to 
decompose the exponential in terms of the spherical harmonics
(eq. \ref{delta2harm}).
Further integrating over the azimuthal angles $\phi_{\ell_1}$,
$\phi_{\ell_2}$, $\phi_{\ell_3}$, collapses the sum to
\be
\langle a_{\ell_1 m_1}^s a_{\ell_2 m_2}^s a_{\ell_3 m_3}^s\rangle\approx
B(\ell_1,\ell_2,\ell_3) \int d^2x W(\bx)^3
\,Y_{\ell_1 m_1}^*(\bx)Y_{\ell_2 m_2}^*(\bx)
Y_{\ell_3 m_3}^*(\bx).
\ee
Comparing this to equation (\ref{incomp_bsp}) we obtain the desired
relation (eq. \ref{flat_approx})
\be
  b_{\ell_1\ell_2\ell_3}\approx
  B(\ell_1,\ell_2,\ell_3).
\ee

\section{Best Cubic estimator: All sky}
\label{cub_gen}

In this appendix we will calculate the optimal
estimator for the reduced bispectrum $b_{\ell_1\ell_2\ell_3}$
in the full sky setup;
\citet{Hea98} calculated the optimal estimator for the third
moment, but clearly it contains no more information for a
statistically isotropic field.

Again, we consider quantities, $y_\alpha$, cubic in the 
$\Delta T_i$
\begin{equation}
y_\alpha = \sum_{{\rm pixels}\ ijk} E^\alpha_{ijk}\ \Delta T_i 
\Delta T_j \Delta T_k,
\label{ap:y_def}
\end{equation}
with $\alpha=\{\ell_1,\ell_2,\ell_3\}$.
To obtain the mean of $y_\alpha$ we have to consider the
3-point correlation function,
which may be written in terms of the reduced bispectrum as follows:
\begin{equation}
\mu_{ijk}\equiv \langle \Delta T_i \Delta T_j \Delta T_k\rangle =
\sum_{\ell_1\,\ell_2\,\ell_3} b_{\ell_1\ell_2\ell_3}
Q_{ijk}^\alpha
\end{equation}
where
\begin{equation}
Q_{ijk}^\alpha = W_{\ell_1}W_{\ell_2}W_{\ell_3}
\sum_{m_1\,m_2\,m_3} 
{\cal G}_{\ell_1\ell_2\ell_3}^{m_1m_2m_3}
Y_{\ell_1 m_1}(\Omega_i)Y_{\ell_2 m_2}(\Omega_j)Y_{\ell_3 m_3}(\Omega_k).
\label{Qijk}
\end{equation}
We have assumed that the noise has a zero 3-point function.  If it
is known and non-zero, it may be added.  The effect of
beam-smearing is taken into account through window
functions, $W_\ell$, multiplying the $a_{\ell m}$ and hence affecting our
estimate of $b_{\ell_1\ell_2\ell_3}$.
Note that \citet{GM00} made some progress in defining an
estimator for the bispectrum, but present final results only for
all-sky coverage.

The remainder of the analysis follows closely that in
\citet{Hea98}. We minimize the variance of $y$, which involves
the 6-point function. The ensemble average of $y$ is
\begin{equation}
\langle y_\alpha \rangle = \sum_{\alpha'ijk} b_{\alpha'}
Q^{\alpha'}_{ijk} E^{\alpha}_{ijk}.
\end{equation}
The covariance between the $y$s is $C_{\alpha\alpha'}\equiv
\langle y_\alpha y_{\alpha'}\rangle - \langle
y_\alpha\rangle\langle y_{\alpha'}\rangle$ which we obtain from
the triplet data covariance matrix:
\begin{equation}
\langle \Delta T_i \Delta T_j \Delta T_k \Delta T_{i'} \Delta T_{j'} 
\Delta T_{k'}\rangle - \langle \Delta T_i \Delta T_j
\Delta T_k\rangle\langle \Delta T_{i'} \Delta T_{j'} \Delta T_{k'}\rangle.
\end{equation}
We again assume that the field is close to a
Gaussian, and approximate the covariance matrix by the covariance
matrix for a Gaussian field with the same power spectrum.  This
assumes that the bispectrum is small compared with the cosmic
variance, and also assumes that the connected 4-point function is
small. The method is therefore optimal for testing the hypothesis
that the field is Gaussian.  For non-Gaussian fields it should be
an accurate estimator provided the bispectrum is small compared
with the cosmic r.m.s., which is the case for most models
considered (an exception is the curvaton model of \citealt{LW2002}, where
the amplitude of the bispectrum is related to other parameters and may be large).

In the Gaussian approximation, $\langle \Delta T_i \Delta T_j \Delta T_k\rangle=0$, and we use
Wick's theorem to write
\begin{eqnarray}
\langle \Delta T_i \Delta T_j \Delta T_k \Delta T_{i'} \Delta T_{j'} 
\Delta T_{k'}\rangle &  = &
\xi_{ij}\xi_{ki'}
\xi_{j'k'}
\nn
& & {\rm\ +  \ permutations\ (15\ terms).}
\end{eqnarray}
where we have defined the 2-point function of the temperature
field:
\begin{equation}
\xi_{ij} \equiv \langle \Delta T_i \Delta T_j\rangle = \sum_\ell {2\ell+1\over
4\pi}\, C_\ell P_\ell (\cos\gamma_{ij})W_\ell^2 + N_{ij}
\end{equation}
$N_{ij}$ is the noise covariance matrix, and $C_\ell = \langle
|a_{\ell m}|^2\rangle$ is the angular power spectrum.  Repeating
exactly the analysis in section 2 of \citet{Hea98}, we obtain the
weights for the optimal estimator for the reduced
bispectrum, by minimizing the error on $y_\alpha$.  The results
only are quoted here.  The optimal weights are
\begin{equation}
E^\alpha_{ijk} = {1\over
6}\xi_{ii'}^{-1}\xi_{jj'}^{-1}\xi_{kk'}^{-1}Q_{i'j'k'}^\alpha
 - {1\over 2(2+3n)}\,\xi_{ii'}^{-1}\xi_{jk}^{-1}\xi_{j'k'}^{-1}
Q_{i'j'k'}^\alpha
\label{ap:E}
\end{equation}
where the summation convention is assumed and $n$ is the number
of pixels used. 
If we plug this expression into equation (\ref{ap:y_def}) and assume the 
second piece (with the factor
${1\over 2(2+3n)}$) is negligible, we obtain
\bea
\lefteqn{y_{\ell_1\ell_2\ell_3} = {1\over 6}W_{\ell_1}W_{\ell_2}W_{\ell_3}
\sum_{m_1\,m_2\,m_3}
{\cal G}_{\ell_1\ell_2\ell_3}^{m_1m_2m_3}\,z_i\,z_j\,z_k\,
Y_{\ell_1 m_1}(\Omega_i)Y_{\ell_2 m_2}(\Omega_j)Y_{\ell_3 m_3}(\Omega_k)}
\nn & &
={1\over 6}W_{\ell_1}W_{\ell_2}W_{\ell_3}
\int d\Omega\, e_{\ell_1}(\Omega)\,e_{\ell_2}(\Omega)\,e_{\ell_3}(\Omega),
\eea
with $z_i=\xi_{ii'}^{-1}\Delta T_{i'}$ and 
\be
e_{\ell}(\Omega)=\sum_m z_i\,Y_{\ell m}(\Omega_i)Y_{\ell m}(\Omega).
\ee
We used equation (\ref{gauntint}) for the Gaunt integral. 
The sum $z_i Y_{\ell m}(\Omega_i)$ over the pixels 
is basically the harmonic transform of the map $z_i$.
The actual estimator for the reduced bispectrum is
\begin{equation}
\hat b_\alpha = F_{\alpha \alpha'}^{-1}y_{\alpha'}
\end{equation}
and the Fisher matrix is then
\begin{eqnarray}
F_{\alpha \alpha'} & = &{\mat C}_{ijki'j'k'}^{-1} Q_{ijk}^{\alpha}
Q_{i'j'k'}^ {\alpha'}\nn & = &
\left[\xi_{ii'}\left(6\xi_{jj'}\xi_{kk'}+9\xi_{jk}\xi_{j'k'}\right)
\right]^{-1}Q_{ijk}^{\alpha} Q_{i'j'k'}^{\alpha'} \label{Ftrip}
\end{eqnarray}
The proof that this weighting scheme is lossless (i.e. $y_\alpha$
contains as much information as the entire pixel dataset) follows
by the Fisher matrix approach in \citet{Hea98}. The method is lossless in the sense
that the Fisher matrices coincide, so the likelihood surfaces are locally identical.

\section{Best cubic estimator: flat case}
\label{flat_estimat}

We will now derive the precise steps to obtain the best cubic
estimator in the flat-sky approximation.
We need first to calculate the 3-point correlation function of the
temperature fluctuations:
\bean
\lefteqn{\langle \Delta T_s(\bx_i) \Delta T_s(\bx_j)
\Delta T_s(\bx_k) \rangle = W(\bx_i)\,W(\bx_j)\,W(\bx_k)} \\ & &
\times\int\frac{d^2k_1}{(2\pi)^2}\int\frac{d^2k_2}{(2\pi)^2}
\,B\left(k_1,k_2,|\bk_1+\bk_2|\right)\,\tilde{\cal B}(k_1)\,\tilde{\cal B}(k_2)
\,\tilde{\cal B}(|\bk_1+\bk_2|)
\,e^{i\bk_1\cdot\bx_i}\,e^{i\bk_2\cdot\bx_j}
\,e^{-i(\bk_1+\bk_2)\cdot\bx_k}.
\eean
The window function $W(\bx_i)$, will just correspond to 0s and 1s, so as to define the field
of observation, and taking into account the rotational symmetry of the bispectrum:
\bean
\lefteqn{\langle \Delta T_s(\bx_i) \Delta T_s(\bx_j)
\Delta T_s(\bx_k) \rangle = 
\int_{0}^{\infty}dk_1\int_{0}^{\infty}dk_2\int_{0}^{\pi}d\theta\,
\frac{2\,k_1\,k_2}{(2\pi)^4}\,B(k_1,k_2,|\bk_1+\bk_2|)} \\ & & 
\times\tilde{\cal B}(k_1)\,\tilde{\cal B}(k_2)
\,\tilde{\cal B}(|\bk_1+\bk_2|)\,
\int_{0}^{\pi}d\theta_1\,{\cal R}e\{
G_{ijk}(k_1,k_2,\theta,\theta_1)+G_{ijk}(k_1,k_2,-\theta,\theta_1)\}
\eean
where,
\be
\label{ap:Gdef}
G_{ijk}(k_1,k_2,\theta,\theta_1) = 
e^{i\bk_1\cdot\bx_i}\,e^{i\bk_2\cdot\bx_j}\,e^{-i\bk_1\cdot\bx_k}\,
e^{-i\bk_2\cdot\bx_k}
\ee
and $\bk_1=(k_1\cos\theta_1,k_1\sin\theta_1)$, 
$\bk_2=(k_2\cos(\theta_1+\theta),k_2\sin(\theta_1+\theta))$.
Let us now do a simple variable transformation,
$\ell_1=k_1,\, \ell_2=k_2$, $\ell_3=\sqrt{k_1^2+k_2^2+2\,k_1k_2\cos(\theta)}$
so that $G_{ijk}(k_1,k_2,\theta,\theta_1)+ G_{ijk}(k_1,k_2,-\theta,\theta_1)\
\rightarrow \tilde{G}_{ijk}(\ell_1,\ell_2,\ell_3,\theta_1)$.
Using the permutation symmetry of the bispectrum, we finally get,
\bea
\lefteqn{\langle \Delta T_s(\bx_i) \Delta T_s(\bx_j)
\Delta T_s(\bx_k) \rangle =
\int_{0}^{\infty}d\ell_1\int_{\frac{\ell_1}{2}}^{\ell_1}d\ell_2
\int_{\ell_1-\ell_2}^{\ell_2}d\ell_3\,B(\ell_1,\ell_2,\ell_3)}\nn & &
\times\tilde{\cal B}(\ell_1)\tilde{\cal B}(\ell_2)\tilde{\cal B}(\ell_3)
\sum_{{\rm perm.}\atop\{\ell_1,\ell_2,\ell_3\}}\int_{0}^{\pi}d\theta_1\,
{\cal R}e\{R_{ijk}(\ell_1,\ell_2,\ell_3,\theta_1)\},
\eea
with, 
\be
\label{ap:Rdef}
R_{ijk}(\ell_1,\ell_2,\ell_3,\theta_1)=
\frac{4\,\ell_1\ell_2\ell_3}{(2\pi)^4\sqrt{(2\ell_1\ell_2)^2-(\ell_3^2-\ell_1^2-\ell_2^2)^2}}\,
\tilde{G}_{ijk}(\ell_1,\ell_2,\ell_3,\theta_1).
\ee
Looking back at equation (\ref{3-point}) it is possible to read off exactly
what should be the expression for $Q_{ijk}(\ell_1,\ell_2,\ell_3)$:
\be
Q_{ijk}(\ell_1,\ell_2,\ell_3)=
\tilde{\cal B}(\ell_1)\tilde{\cal B}(\ell_2)\tilde{\cal B}(\ell_3)
\sum_{{\rm perm.}\atop\{\ell_1,\ell_2,\ell_3\}}\int_{0}^{\pi}d\theta_1\,
{\cal R}e\{R_{ijk}(\ell_1,\ell_2,\ell_3,\theta_1)\},
\ee
Then the expression for $y_\alpha$, or 
$y(\ell_1,\ell_2,\ell_3)$ in the flat case, will be
\bea
\label{ap:flat_ydef}
\lefteqn{y(\ell_1,\ell_2,\ell_3) = \tilde{\cal B}(\ell_1)\tilde{\cal B}(\ell_2)\tilde{\cal B}(\ell_3)\,
{1\over 6}\sum_{\rm perm.}\int_{0}^{\pi}d\theta_1
{\cal R}e\{R_{i'j'k'}(\ell_1,\ell_2,\ell_3,\theta_1)} \nonumber \\ & &
\times\xi_{ii'}^{-1}\left[
\xi_{jj'}^{-1}\xi_{kk'}^{-1}
 - {3\over 2+3n}\,\xi_{jk}^{-1}\xi_{j'k'}^{-1}\right]\,
\Delta T_s(\bx_i)\Delta T_s(\bx_j)\Delta T_s(\bx_k)\}
\eea
The ensemble average of y gives
\be
\label{ap:ytobsp}
\langle y(\ell_1,\ell_2,\ell_3)\rangle=
\int_{0}^{\infty}d\ell_1'\int_{\frac{\ell_1'}{2}}^{\ell_1'}d\ell_2'
\int_{\ell_1'-\ell_2'}^{\ell_2'}d\ell_3'
\,F(\ell_1,\ell_2,\ell_3,\ell_1',\ell_2',\ell_3')\,B(\ell_1',\ell_2',\ell_3')
\ee
With
\bea
\label{ap:fisher_def}
\lefteqn{F(\ell_1,\ell_2,\ell_3,\ell_1',\ell_2',\ell_3')=
\langle y(\ell_1,\ell_2,\ell_3)\,y(\ell_1',\ell_2',\ell_3')\rangle}
\nn & &
=\tilde{\cal B}(\ell_1)\tilde{\cal B}(\ell_2)\tilde{\cal B}(\ell_3)\,
{1\over 6}\sum_{\rm perm.}\int_{0}^{\pi}d\theta_1
{\cal R}e\{R_{ijk}(\ell_1,\ell_2,\ell_3,\theta_1)\}
\nn & &
\times\xi_{ii'}^{-1}\left[
\xi_{jj'}^{-1}\xi_{kk'}^{-1}
 - {3\over 2+3n}\,\xi_{jk}^{-1}\xi_{j'k'}^{-1}\right]\,
\tilde{\cal B}(\ell'_1)\tilde{\cal B}(\ell'_2)\tilde{\cal B}(\ell'_3)
\nn & &
\times\sum_{\rm perm.}\int_{0}^{\pi}d\theta'_1
{\cal R}e\{R_{i'j'k'}(\ell'_1,\ell'_2,\ell'_3,\theta'_1)\}
\eea
In the paper we used the following estimator for the bispectrum:
\be
\hat{B}(\ell_1,\ell_2,\ell_3)=\frac{y(\ell_1,\ell_2,\ell_3)}
{M(\ell_1,\ell_2,\ell_3)},
\ee
where
\bea
\label{ap:M}
\lefteqn{M(\ell_1,\ell_2,\ell_3) = 
\int_{0}^{\infty}d\ell_1'\int_{\frac{\ell_1'}{2}}^{\ell_1'}d\ell_2'
\int_{\ell_1'-\ell_2'}^{\ell_2'}d\ell_3'
\,F(\ell_1,\ell_2,\ell_3,\ell_1',\ell_2',\ell_3')}
\nn & &
={1\over 6}\int d^2x \,{\cal B}(|\bx_{i'}-\bx|){\cal B}(|\bx_{j'}-\bx|)
{\cal B}(|\bx_{k'}-\bx|)
\nn & &
\times\sum_{perm.}\int_{0}^{\pi}d\theta_1
{\cal R}e\{R_{i'j'k'}(\ell_1,\ell_2,\ell_3,\theta_1)\}
\nn & &
\times\xi_{ii'}^{-1}\left[
\xi_{jj'}^{-1}\xi_{kk'}^{-1}
 - {3\over 2+3N}\,\xi_{jk}^{-1}\xi_{j'k'}^{-1}\right]
\tilde{\cal B}(\ell_1)\tilde{\cal B}(\ell_2)\tilde{\cal B}(\ell_3).
\eea
If we approximate the beam window function by a Gaussian,
\be
{\cal B}(|\bx|)=A\,e^{-\frac{x^2}{2\sigma^2}},
\ee
the above integral will simplify, giving
\bea
\label{ap:M_G}
\lefteqn{M(\ell_1,\ell_2,\ell_3) = 
A^3\,\frac{2\pi\sigma^2}{3}\,e^{-{1\over{6\sigma^2}}\left[
|\bx_{i'}-\bx_{j'}|^2+|\bx_{i'}-\bx_{k'}|^2+
|\bx_{j'}-\bx_{k'}|^2\right]}}
\nn & &
\times{1\over 6}\sum_{perm.}\int_{0}^{\pi}d\theta_1
{\cal R}e\{R_{i'j'k'}(\ell_1,\ell_2,\ell_3,\theta_1)\}
\nn & &
\times\xi_{ii'}^{-1}\left[
\xi_{jj'}^{-1}\xi_{kk'}^{-1}
 - {3\over 2+3N}\,\xi_{jk}^{-1}\xi_{j'k'}^{-1}\right]
\tilde{\cal B}(\ell_1)\tilde{\cal B}(\ell_2)\tilde{\cal B}(\ell_3).
\eea
Note that the beam window function ${\cal B}(|\bx|)$ quickly
becomes negligible for large $|\bx|$. Basically this means
the sum above does not have to be done for all ${i'\,j'\,k'}$,
so that we end up with a sum of approximately order
$n$ instead of the original $n^3$.

\end{document}